\documentclass[%
reprint, 
prx,
amsmath,amssymb,
aps, 
]{revtex4-2}

\usepackage{graphicx}
\usepackage{dcolumn}
\usepackage{bm}
\usepackage{dcolumn}
\usepackage{bm}
\usepackage{amssymb}
\usepackage{amsmath}
\usepackage{amsthm}
\usepackage{chemarrow}
\usepackage{multirow}
\usepackage{color}
\usepackage{bigstrut}
\usepackage{epstopdf}
\usepackage{bbold}
\usepackage{youngtab}
\usepackage{mathtools}
\usepackage{setspace}
\usepackage{comment}

\usepackage{mathrsfs}

\usepackage{MnSymbol}

\usepackage{algorithm}
\usepackage{algorithmicx}
\usepackage{algpseudocode}
\usepackage{caption}

\usepackage{booktabs}

\usepackage{youngtab}

\theoremstyle{definition}

\theoremstyle{remark}

\newcommand{\lr}[1]{\left({#1}\right)}
\newcommand{\ls}[1]{\left[{#1}\right]}
\newcommand{\lc}[1]{\left\{{#1}\right\}}

\newcommand{\x}{\sigma_x}
\newcommand{\y}{\sigma_y}
\newcommand{\z}{\sigma_z}

\emergencystretch=\maxdimen
\begin{document}
	\title{Engineering Higher-order Effective Hamiltonians}
\preprint{APS/123-QED}

\author{Jiahui Chen$^{1,2}$}
\email{j562chen@uwaterloo.ca}
\author{David Cory$^{1,2}$}%

\affiliation{%
	$^{1}$Institute for Quantum Computing, Waterloo, Ontario N2L 3G1, Canada\\
	$^{2}$Department of Chemistry, University of Waterloo, Waterloo, Ontario N2L 3G1, Canada
}%


\date{\today}
\begin{abstract}
	Advancing quantum technologies requires precise and robust coherent control of quantum systems. Robust higher-order Hamiltonian engineering is essential for high-precision control and for accessing effective dynamics absent at zeroth order.
	Here, we introduce a systematic methodology for achieving the precision, robustness, and complexity required for quantum control through the engineering of higher-order processes and effective Hamiltonians. We identify the minimal subspace of achievable effective Hamiltonian at each order and provide universal cost functions for achieving desired targets. Examples include robust sequences for decoupling, three-body interactions and detuning/interaction correlations.
\end{abstract}
\maketitle
	\section{Introduction}
	Quantum technologies rely on coherent manipulation of quantum systems to achieve advantages in sensing \cite{giovannetti2006quantum,degen2017quantum}, secure communication \cite{bennett1992quantum}, computation \cite{deutsch1985quantum}, and physics simulation \cite{feynman2018simulating,somaroo1999quantum,lloyd1996universal,tseng1999quantum}. Efficient, precise, and robust control strategies are central to quantum control \cite{brif2010control,dong2010quantum,hincks2020generating,cappellaro2006principles,cappellaro2007dynamics,borneman2010application,pravia2003robust,hodges2008universal,glaser2015training,d2021introduction}, as they directly determine the performance, reliability, and scalability of quantum systems.
	Effective Hamiltonian engineering using Average Hamiltonian Theory (AHT) \cite{haeberlen1968coherent,haeberlen2012high} has proven to be a powerful tool for designing target unitary propagators across diverse systems \cite{waugh1968approach,rhim1971time,sorensen2000entanglement,porras2004effective,magesan2020effective,caldwell2018parametrically,de2010universal,choi2020robust}. It provides a systematic framework for handling uncertainties and control imperfections \cite{fortunato2002design,viola1999dynamical,khodjasteh2009dynamically,levitt1986composite,wimperis1994broadband,ezzell2023dynamical,haas2019engineering}.
	
	Consider a system with $L$ control channels and total Hamiltonian
	\begin{equation}
		H_{\mathrm{tot}}(t;z)=\sum_{l=1}^{L}H_c^{(l)}\!\big(\vec B_l(t;z)\big)+H_{\mathrm{int}}\!\big(\{\eta_{\mathrm{int}}^i(z)\}\big),
		\label{totH}
	\end{equation}
	where the field actually experienced by the qubits is
	\begin{equation}
		\vec B_l(t;z)=f_l\!\big(\{a_{l',k}(t)\},\lc{\mu_{l,j}(z)}\big).
	\end{equation}
	 For each channel $l$, the controller outputs a programmed waveform given by the control parameters $\lc{a_{l,k}(t)}$,
	which serves as the input to an input--output model that maps it to this field. The function $f_l$ may be specified by a transfer function, a nonlinear distortion model, or an equivalent state-space description. Here $\lc{\mu_{l,j}(z)}$ collect all model parameters of channel $l$ (gain/phase factors, timing offsets, bandwidth/phase-lag coefficients, reactance, cross-talk coefficients, and nonlinear distortion parameters), and the dependence on $\{a_{l',k}(t)\}$ for $l'\neq l$ captures cross-talk between channels. $z$ labels a realization in an ensemble (e.g., spatial position in an inhomogeneous sample or an experimental realization), so that both the model parameters $\{\mu_{l,j}(z)\}$ and the internal parameters $\{\eta_{\mathrm{int}}^i(z)\}$ can vary across the ensemble. $\{\eta_{\mathrm{int}}^i\}$ include coupling strengths and detunings/offsets. We treat $(\{\mu_{l,j}\},\{\eta_{\mathrm{int}}^i\})$ as classical uncertain variables characterized by a distribution $p(z)$.
	
	A common figure of merit (FoM) is the expected gate overlap with a target unitary $U_0$,
	\begin{equation}
		\begin{split}
		\text{FoM}&=\int dz\, p(z)\,
		\frac{\big|\mathrm{Tr}\!\big(U(T_{\mathrm{seq}};z)^\dagger U_0\big)\big|}{\mathrm{Tr}(U_0^\dagger U_0)}\\
		&=\int dz~p(z)\mathcal F(U(T_{\mathrm{seq}};z)),
		\end{split}
	\end{equation}
	where
	\begin{equation}
		U(T_\text{seq})=\mathcal T\exp\lr{-i\int_0^{T_\text{seq}}H_\text{tot}(t)dt}.
		\label{ute}
	\end{equation} 
	Other choices of $\mathcal F$ include average gate fidelity, state overlap with a target state, and Fisher information, etc.
	The goal of quantum control is to design the time dependence of $\lc{a_{l,k}(t)}$ to maximize the FoM.
	 
	Equivalently, one can always evaluate $U(T_\text{seq})$ through an effective Hamiltonian $H_\text{eff}$:
	\begin{equation}
		U(T_\text{seq})=\exp(-i H_\text{eff} T_\text{seq}).
	\end{equation}
	A simple, effective strategy to maximize the FoM is then to design for a target effective Hamiltonian that achieves high $\mathcal F$ regardless of the values of $z$.
	
	AHT provides a general framework for designing precise target effective Hamiltonian while suppressing its sensitivity to variations in model and internal parameters.
	It typically evaluates $H_\text{eff}$ perturbatively using the Magnus expansion.
	 The zeroth-order average Hamiltonian \cite{Note1}, $\bar H^{(0)}$, provides the simplest and typically the most efficient implementation when the target is achievable at this order. 
	
	However, engineering higher-order terms can be useful in the following situations:
	\begin{enumerate}
		\item When the target control is achieved at zeroth order,  minimizing higher-order terms ($\bar H^{(r-1)}=0, \forall r>1$), especially components that commute with the target effective \cite{haeberlen2012high,cory1990multiple,chen2025engineering}, is important for achieving high precision and robustness (e.g., for decoupling \cite{mansfield1971symmetrized,rhim1973analysis,burum1979analysis,cory1990time} and gate optimization \cite{zhao2025higher,green2012high,brown2004arbitrarily}). 
		\item When target unitary includes multi-qubit correlations absent in the native Hamiltonians, it may be more convenient to achieve the target in higher orders. These correlations often emerge through the nested commutators of the toggling-frame Hamiltonian that make up the higher-order contributions to the average Hamiltonian.
		\item When one wishes to engineer nonlinear dependence on Hamiltonian parameters that are not available at zeroth order. 
	\end{enumerate}
	
	Engineering higher-order effective Hamiltonians is challenging. Higher-order contributions involve nested commutators, which generate operators spanning a high-dimensional operator space. In addition, direct numerical evaluation of the Magnus expansion can be inconvenient in the presence of uncertain parameters or infinite-dimensional Hamiltonians.
	
	Symmetry-based approaches are often effective for engineering higher-order terms. For example, first-order corrections can be eliminated by symmetrizing a control sequence \cite{haeberlen1968coherent,mansfield1971symmetrized,burum1981magnus}, a technique that is widely used. Selective properties of other symmetry classes have also been studied extensively and used to design robust sequences \cite{eden1999pulse,levitt2008symmetry,carravetta2000symmetry,levitt2007symmetry}. Such symmetry-based solutions are independent of Hamiltonian parameter values and enable scalable designs that are insensitive to system dimension. 
	
	However, existing symmetry-based solutions often discard Hamiltonian-specific information and can be overly restrictive to implement. Moreover, general symmetry conditions are primarily designed to eliminate higher-order terms. In contrast, engineering nonzero higher-order effects remains strongly case-dependent and is not easily transferable across systems.
	
	There have been efforts toward engineering higher-order effective Hamiltonians. Higher-order robustness to Rabi field inhomogeneity and variations in qubit detunings has been explored extensively in the design of composite pulses \cite{wimperis1994broadband,brown2004arbitrarily}. W. Warren \textit{et al.} showed the importance of managing higher-order terms when selectively exciting $n-$quantum coherences \cite{warren1979selective,warren1980theory,warren1980high} and applying shaped pulses \cite{warren1984effects}. Higher-order terms can generate multi-body (multi-quantum) dynamics and are crucial in recoupling \cite{brinkmann2004second,eden2005triple,eden2002order,ernst1996second}.
	
	Recently, effective multi-body interactions have been studied on different platforms. X. Peng \textit{et al.} \cite{peng2009quantum} studied two- and three-body interactions using a small NMR quantum processor.
	An engineered three-body interaction of spin chirality through first-order effects using periodic drive has been demonstrated on superconducting qubits \cite{liu2020synthesizing}. Tunable three-body interactions through virtual higher-order transitions have also been demonstrated \cite{menke2022demonstration}. B. Andrade \textit{et al.} \cite{andrade2022engineering} proposed a method to achieve three-body interactions with trapped ions. O. Katz \textit{et al.} \cite{katz2023demonstration} experimentally demonstrated how to achieve three- and four-body interactions between trapped-ion spins. C. Luo \textit{et al.} \cite{luo2025realization} realized three- and four-body interactions between atoms in a cavity by suppressing zeroth-order two-body interactions through symmetry. There are also theoretical proposals for engineering various types of multi-body interaction in different systems \cite{jordan2008perturbative,cao2014perturbative,chancellor2017circuit,leib2016transmon,katz2022n,katz2023programmable,honer2010collective}.
	Y. Xu \textit{et al.} \cite{xu2025perturbative} discussed higher-order Floquet engineering of a Harmonic oscillator.
	
	J. Choi \textit{et al.} \cite{choi2020robust} proposed a framework for designing decoupling sequences in the context of anisotropic Heisenberg interactions using collective $\pi/2$ rotations and can be generalized to engineering qudit systems \cite{zhou2024robust}. 
	Minimization of higher-order corrections have also been analyzed  \cite{zhou2023robust,tyler2023higher}.
	
	Despite growing interest in engineering higher-order dynamics, many existing approaches remain highly system- and task-specific, relying on substantial expert intuition and on features of particular platforms or Hamiltonians that do not readily generalize. To our knowledge, a general, system-agnostic framework that enables systematic higher-order engineering in the presence of uncertainties remains unavailable.
	
	A central challenge in Hamiltonian engineering is the absence of a framework that simultaneously exploits Hamiltonian-specific information, supports numerical optimization, and accommodates parameter uncertainty. Conventional use of AHT captures how Hamiltonians transform under control but relies heavily on intuition and becomes impractical for large operator spaces or multiple objectives, whereas numerical methods for evaluating the Magnus expansion or optimal control typically require Hamiltonian parameters (e.g., coupling strengths and qubit detunings) to be fixed \cite{fortunato2002design,khaneja2005optimal,caneva2011chopped}. What is missing is a bridge between these approaches that enables numerical search while allowing selected parameters to enter quantitatively, and enforcing by construction either retention of or insensitivity to the remaining parameters in the engineered Hamiltonian.
	
	J. Chen \textit{et al.} \cite{chen2025engineering} introduced a general framework for engineering zeroth-order average Hamiltonians and minimizing higher-order corrections including all cross terms in arbitrary Hamiltonian systems. However, the higher-order correction conditions in \cite{chen2025engineering} discard the Hamiltonian information beyond the minimal error space, therefore are generally overly restrictive. Moreover, Ref. \cite{chen2025engineering} does not address engineering nonzero higher-order dynamics.
	
	In order to build a general framework for engineering higher-order dynamics, we still need
	\begin{enumerate}
		\item {\bf Controllability}: A general method for characterizing the minimal space of achievable higher-order effective Hamiltonians at each order;
		\item {\bf Higher-order corrections}: A method to find necessary and sufficient conditions for higher-order corrections;
		\item {\bf Engineer nonzero higher-order terms}: Numerical objectives for engineering desired scalable higher-order terms.
	\end{enumerate}
	
This work presents a systematic method based on the framework in \cite{chen2025engineering}, applicable to arbitrary systems, for determining the controllability of higher-order terms and formulating numerical objectives for engineering them. In doing so, it enables the design of control sequences with greater precision, robustness and efficiency than conventional methods \cite{cory1990multiple,cory1990time,burum1981magnus} by accounting for higher-order corrections, while also providing access to higher-order and nonlinear effects when the desired target cannot be realized at zeroth order.
The framework enables design of parameter-retaining solutions that remain valid regardless of the values of parameters defined in a parameter graph, while it reduces to exact numerical Magnus expansion when all parameters are exact. Using the parameter graphs, we can find the minimal subspace of achievable average Hamiltonians at each order and provide numerical cost functions for engineering arbitrary achievable higher-order targets. As an application, we show how to engineer higher-order terms on qubit networks with collective control, and provide examples in a dipole-coupled ensemble with all-to-all interactions in the presence of Rabi field variations and qubit detunings. We demonstrate designing a robust decoupling sequence that greatly outperforms state-of-the-art alternatives \cite{cory1990time,zhou2023robust,peng2022deep}; a robust, scalable, parameter-retaining solution for engineering three-body interactions; and creating a detuning/interaction cross term that enables measurement of their correlation.
	
The rest of the paper is organized as follows. Sec. \ref{sec2} reviews the framework for zeroth-order Hamiltonian engineering \cite{chen2025engineering}. Sec. \ref{sec3} gives the general method for engineering higher-order terms. Sec. \ref{secfun} gives the total cost function. Engineering qubit networks is discussed in Sec. \ref{QN1}, followed by examples (Sec. \ref{secexa}), a discussion and a conclusion (Secs. \ref{secdis} and \ref{secon}). 	
	\section{Engineering zeroth-order average Hamiltonians}\label{sec2}
	Here, we briefly review the procedure for engineering zeroth-order average Hamiltonians introduced in \cite{chen2025engineering}, which provides the foundation for engineering higher-order effective Hamiltonians.
	The total Hamiltonian $H_\text{tot}$ is partitioned into primary and perturbative parts:
	\begin{equation}
		H_\text{tot}(t)=H_\text{pri}(t)+H_\text{pert}(t).
		\label{par}
	\end{equation}
	The toggling frame is the interaction frame of the primary Hamiltonian $H_\text{pri}(t)$.
	
	The toggling-frame Hamiltonian of $H_\text{pert}(t)$ is
	\begin{equation}
		H_\text{tog}(t)=U_\text{pri}^\dagger(t)H_\text{pert}(t)U_\text{pri}(t),
	\end{equation}
	where the primary unitary $U_\text{pri}(t)$ is 
	\begin{equation}
		U_\text{pri}(t)=\mathcal T\exp\left(-i\int_0^{t}H_\text{pri}(t')dt'\right),
	\end{equation}
	and the perturbative unitary is
	\begin{equation}
		\begin{split}
			U_\text{pert}({T_\text{seq}})&=\mathcal T\exp\left(-i\int_0^{T_\text{seq}}H_\text{tog}(t)dt\right)\\
			&=\exp\left(-i H_\text{eff} T_\text{seq}\right).
		\end{split}
	\end{equation}
	The total propagator is 
	\begin{equation}
		\begin{split}
			U({T_\text{seq}})=U_\text{pri}({T_\text{seq}})U_\text{pert}(T_\text{seq}),
		\end{split}
	\end{equation}
	where $U_\text{pri}$ and $U_\text{pert}$ can be evaluated through numerical integration and perturbative expansion, respectively. AHT uses the Magnus expansion to evaluate the effective Hamiltonian $H_\text{eff}=\sum_{r=1}^\infty \bar H^{(r-1)}$ with
	\begin{equation}
		\begin{split}
			&\bar H^{(0)}=\frac{1}{T_\text{seq}}\int_0^{T_\text{seq}}dt_1 H_\text{tog}(t_1),\\
			&\bar H^{(1)}=-\frac{i}{2T_\text{seq}}\int_0^{T_\text{seq}}dt_1\int_0^{t_1}dt_2[ H_\text{tog}(t_1),H_\text{tog}(t_2)],\\
			&\bar H^{(2)}=\\
			&-\frac{1}{6T_\text{seq}}\int_0^{T_\text{seq}}dt_1\int_0^{t_1}dt_2\int_0^{t_2}dt_3\\
			&~~~~~~~~~~~~~~\{[H_\text{tog}(t_1),[H_\text{tog}(t_2),H_\text{tog}(t_3)]]\\
			&~~~~~~~~~~~~~~~~+[H_\text{tog}(t_3),[H_\text{tog}(t_2),H_\text{tog}(t_1)]]\},\\
			&\cdots.
		\end{split}
		\label{magnus}
	\end{equation}
	The minimal subspace of $H_\text{tog}(t)$ is
	\begin{equation}
		\mathscr C(\mathbf g_\text{pri},H_\text{pert})=\text{span}_{\mathbb R}\lc{[g_1,\cdots[g_L,H_\text{pert}]\cdots]|g_l\in\mathbf{g}_\text{pri},L\ge0},
	\end{equation}
	where $\mathbf g_\text{pri}$ is the minimal Lie algebra containing $H_\text{pri}(t)$. The set of achievable zeroth-order average Hamiltonian is a convex set in $\mathscr C(\mathbf g_\text{pri},H_\text{pert})$ and can be characterized through linear programming. A systematic method for achieving robustness against imperfect controls is in \cite{chen2025engineering}. 
	
	Given an orthonormal basis of $\mathscr C(\mathbf g_\text{pri}, H_\text{pert})$:
	\begin{equation}
		\mathcal B(\mathscr C)=\lr{h_1,h_2,\dots,h_{|\mathscr C|}},
	\end{equation}
where $|\mathscr C|$ is the dimension of $\mathscr C(\mathbf g_\text{pri},H_\text{pert})$,
	 the toggling-frame Hamiltonian can be represented as a vector in $\mathscr C(\mathbf g_\text{pri},H_\text{pert})$ as
	\begin{equation}
		|H_\text{tog}(t)\rrangle=\lr{c_1(t),c_2(t),\ldots,c_{|\mathscr C|}(t)}.
	\end{equation}
	Denote the time-ordered integrals of the coefficients $c_i(t)$ ($\mathscr C-$integrals) as
	\begin{equation}
		\bar c_{i_1\cdots i_r}=\int_0^{T_\text{seq}}dt_1\cdots\int_0^{t_{r-1}}dt_rc_{i_1}(t_1)\cdots c_{i_r}(t_r),
		\label{Cint}
	\end{equation}
	a generic condition for $\bar H^{(r-1)}=0$ is
	\begin{equation}
		\bar c_{i_1i_2\cdots i_r}=(-1)^r\bar c_{i_ri_{r-1}\cdots i_1},
		\label{ocom}
	\end{equation}
	for all $i_1,i_2,\ldots,i_r$ except when $i_1=i_2=\cdots=i_r$. Sometimes, it is convenient to decompose $H_\text{pert}$ into several components $H_\text{pert}=\sum_{w=1}^WH_\text{pert}^w$ (e.g., when $H_\text{pert}^w$ transform differently under $e^{\mathbf g_\text{pri}}$ and are subject to different types of variations). Then we can compute $\mathscr C_w(\mathbf g_\text{pri},H_\text{pert}^w)$ for each $w$. Define a composite $\mathscr C$ space as
	\begin{equation}
		\mathscr C_\text{comp}=\bigoplus_{w=1}^W \mathscr C_w(\mathbf g_\text{pri},H_\text{pert}^w),
		\label{compc}
	\end{equation}
	then the composite toggling-frame Hamiltonian is
	\begin{equation}
		\begin{split}
			|H_\text{tog}(t)\rrangle_\text{comp}&=\bigoplus_{w=1}^W|H_\text{tog}(t)^w\rrangle_w\\
			&=(c_1^1(t),\ldots,c_{|\mathscr C_1|}^1,\ldots,c_1^W(t),\ldots,c_{|\mathscr C_W|}^W)\\
			&=(c_1^\text{comp}(t),\ldots,c_{|\mathscr C_\text{comp}|}^\text{comp}(t)).
		\end{split}
	\end{equation}
	The condition for engineering each $H_\text{pert}^w$ to $H_\text{target}^w$ with a scaling factor $s_w$ is
	\begin{equation}
		\begin{split}
			\frac{1}{T_\text{seq}}\int_0^{T_\text{seq}}|H_\text{tog}(t)\rrangle_\text{comp}&=\frac{1}{T_\text{seq}}(\bar c_1^\text{comp}(t),\ldots,\bar c_{|\mathscr C_\text{comp}|}^\text{comp}(t))\\
			&=\bigoplus_{w=1}^W |s_wH_\text{target}^w\rrangle_w.
		\end{split}
		\label{tot0}
	\end{equation}
	The condition for minimizing higher-order corrections, including cross terms between different $H_\text{pert}^w$, is
	\begin{equation}
		\bar c_{i_1\cdots i_r}^\text{comp}(T_\text{seq})-(-1)^{r-1}\bar c_{i_r\cdots i_1}^\text{comp}(T_\text{seq})=0,
		\label{totr}
	\end{equation}
	for all $i_1,\ldots,i_r$ excluding the situations where $i_1=\cdots=i_r$.
	The composite $\mathscr C$-integrals admit exact analytic solutions, expressed as explicit functions of the eigenvalues of the primary Hamiltonian, for piecewise-constant modulations. The package for calculating  composite $\mathscr C$-integrals is provided in \cite{chen2025engineering}. 
	\section{General method for Engineering $\bar H^{(r-1)}$}\label{sec3}
	This section develops a general framework for characterizing the minimal space of achievable effective Hamiltonians and gives the conditions for engineering any target from this space at arbitrary order.
	
	First, consider the case when $W=1$, so, $H_\text{pert}=H_\text{pert}^1$ and $\mathscr C_\text{comp}=\mathscr C(\mathbf g_\text{pri},H_\text{pert})$.
	Expressing the Magnus expansion (Eqs. \eqref{magnus}) in terms of the $\mathscr C-$integrals yields
	\begin{equation}
		\begin{split}
			\bar H^{(r-1)}T_\text{seq}= \sum_{i_1,\ldots ,i_r=1}^{|\mathscr C|}F_{i_1\cdots i_r} \bar c_{i_1\cdots i_r}.
		\end{split}
	\label{magc}
	\end{equation}
	where 
	\begin{equation}
		\begin{split}
			F_{i_1\cdots i_r}
			=\sum_{\pi}(-i)^{r-1}(-1)^{d_b}\frac{d_a!d_b!}{r!}\pi\lr{h_{i_1}h_{i_2}\cdots h_{i_r}},
		\end{split}
	\label{Fle}
	\end{equation}
	where the sum is taken over all possible permutations $\pi$ of $h_{i_1},h_{i_2},\ldots,h_{i_r}$. $d_a$ and $d_b$ are ascents and descents of $\pi$ \cite{agrachev2017shuffle,arnal2018general}. Define the vectorized $\mathscr C-$integrals as
	\begin{equation}
		\vec c^{(r-1)}=(\bar c_{1\cdots1 1},\bar c_{1\cdots1 2},\ldots,\bar c_{|\mathscr C|\cdots|\mathscr C| |\mathscr C|})^T,
	\end{equation} 
	For simplicity, assume that all possible $\vec c^{(r-1)}$ span the space $\mathscr C(\mathbf g_\text{pri},H_\text{pert})^{\otimes r}$, the general situation where this condition may not be met is discussed in Appendix \ref{appen1}. Then $\lc{F_{i_1\cdots i_r}}$ spans the minimal subspace of achievable $\bar H^{(r-1)}$. 
	When $H_\text{pert}$ is finite-dimensional and known exactly, each $F_{i_1\cdots i_r}$ can be calculated exactly. Thus, engineering $\bar H^{(r-1)}$ through Eq. \eqref{magc} is equivalent to using any numerical Magnus-expansion tool. When $H_\text{pert}$ is not finite-dimensional, if all parameters in $H_\text{pert}$ are known exactly, $F_{i_1\cdots i_r}$ typically have analytic forms. Eq. \eqref{magc} gives the analytic form of $\bar H^{(r-1)}$ and engineering a target $\bar H_\text{target}^{(r-1)}$ can be achieved through Eq. \eqref{magc} by evaluating the $\mathscr C-$integrals.
		  
	When $H_\text{pert}$ includes parameters, $\lc{\eta_\text{pert}^m}_{m=1}^M$, that are subject to variations, we seek solutions that hold for arbitrary values of $\lc{\eta_\text{pert}^m}$. Assuming that $H_\text{pert}$ is linear in $\lc{\eta_\text{pert}^m}$, then $\bar H^{(r-1)}$ are polynomials of $\lc{\eta_\text{pert}^m}$ of order up to $r$. For convenience, we absorb the Hamiltonian's units into the $\mathscr C$-integrals so that all coefficients are dimensionless. This is achieved by expressing time and frequency in reciprocal units (e.g., time in $\mu\mathrm{s}$ and frequency in MHz).
	Define a parameter graph $G=\lc{m_1,\ldots,m_{r'}}$ for $m_1,m_2,\ldots,m_{r'}\in\lc{1,2,\ldots,M}$ and $r'=0,1,\ldots,r$ and
	\begin{equation}
		\partial_G=\left.\frac{\partial^{r'}}{\partial \eta_\text{pert}^{m_1}\cdots\partial \eta_\text{pert}^{m_{r'}}}\right|_{\vec\eta_\text{pert}=\vec 0}.
	\end{equation}
Then $\bar H^{(r-1)}$ can be written as 
\begin{equation}
	\bar H^{(r-1)}=\sum_{G}\partial_G\bar H^{(r-1)}\prod_{m\in G}\eta_\text{pert}^m.
\end{equation}
Therefore, for a desired target $H_\text{target}^{(r-1)}$, $\bar H^{(r-1)}=\bar H_\text{target}^{(r-1)}$ for all $\lc{\eta_\text{pert}^m}$ if and only if
\begin{equation}
	\partial_G\bar H^{(r-1)}=\partial_G\bar H^{(r-1)}_\text{target},
	\label{xcom}
\end{equation} 
for all $G$. Substituting Eq. \eqref{xcom} into Eq. \eqref{magc} yields
\begin{equation}
	\sum_{i_1,\ldots ,i_r=1}^{|\mathscr C|}\partial_GF_{i_1\cdots i_r} \bar c_{i_1\cdots i_r}=\partial_G\bar H_\text{target}^{(r-1)}T',
	\label{fcom}
\end{equation}
for all $G$. Here, we use a fixed time $T'$ instead of $T_\text{seq}$ to emphasize that the goal is to achieve a fixed total action $\bar{H}_\text{target}^{(r-1)} T'$ and does not depend on the value of $T_\text{seq}$. A $\bar H_\text{target}^{(r-1)}$ that satisfies Eq. \eqref{fcom} for some $\lc{a_k(t)}$ and $T'>0$ is an achievable effective Hamiltonian at $(r-1)$th order compatible with $G$ ($G-$compatible). It is uniquely represented by $\lc{\partial_G\bar H_\text{target}^{(r-1)}}$. Thus, the minimal space of achievable $G-$compatible $\bar H_\text{target}^{(r-1)}$ is isomorphic to
\begin{equation}
	\mathscr S(G)=\text{span}_{\mathbb R}\lc{\partial_GF_{i_1\cdots i_r}}_{i_1,\ldots,i_r=1}^{|\mathscr C|}.
	\label{gsub}
\end{equation} 
Since $\bar H_\text{target}^{(r-1)}=0$ is in the interior of the set of $G-$compatible $\bar H_\text{target}^{(r-1)}$, any element from $\mathscr S(G)$ is achievable for some $T'>0$. 
Eq. \eqref{fcom} can be formally written as
\begin{equation}
	\mathbf A(G)\cdot \vec c^{(r-1)}=\vec d(G),
	\label{gcon}
\end{equation}
where the columns of $\mathbf A(G)$ are vec$[\partial_GF_{i_1\cdots i_r}]$ and $\vec d(G)=\text{vec}[\partial_G\bar H_\text{target}^{(r-1)}]T'$. The vectorization operator, vec$[\cdot]$, is a map to a finite linear space. This is always possible, because $\mathscr S(G)$ is at most of dimension $|\mathscr C|^r$. Thus, the validity of Eq. \eqref{gcon} does not explicitly depend on the dimension of $H_\text{pert}$. For finite-dimensional $\partial_GF_{i_1\cdots i_r}$, vec$[\cdot]$ can be a simple column stacking. 

If $\mathbf A(G_1)$ and $\mathbf A(G_2)$ are row-equivalent, $\mathbf A(G_1)=T\mathbf A(G_2)$ where $T$ is invertible, then $\vec d(G_2)=T^{-1}\vec d(G_1)$. Thus, in this case, we only need to consider $G_1$ and $G_2$ carries no additional information. $G_1$ and $G_2$ are equivalent to each other.
Achievable $\bar H_\text{target}^{(r-1)}$ compatible with all (inequivalent) $G$ correspond to $\vec d(G)$ so that Eqs. \eqref{gcon} are mutually compatible for all (inequivalent) $G$. Practically, one can choose the targets $\vec d(G)$ for each $G$ one by one. Now assume that valid targets have been chosen for $G_1,\ldots,G_j$, then the valid choices of $\vec d(G_{j+1})$ are 
\begin{equation}
	\vec d(G_{j+1})\in \mathbf A(G_{j+1}) \ls{\vec c_j +\bigcap_{i=1}^j\mathcal N(\mathbf A(G_i))},
	\label{gtar}
\end{equation}
where $\vec c_j$ is a particular solution of the linear system given by Eq. \eqref{gcon} for $G_1,\ldots,G_j$ and $\mathcal N(\mathbf A(G_i))$ is the nullspace of $\mathbf A(G_i)$.

When $H_\text{pert}=\sum_{w=1}^WH_\text{pert}^w$, $W>1$, Eq. \eqref{magc} is expressed in the composite $\mathscr C$ space (Eq. \eqref{compc}) with $F_{i_1\cdots i_r}$ defined under a basis of $\mathscr C_\text{comp}$ and the $\mathscr C-$integrals replaced by the composite $\mathscr C-$integrals. The rest of method (everything that leads to Eq. \eqref{gcon}) for engineering $\bar H^{(r-1)}$ stays the same.

Setting $\vec d(G)=0$ in Eq. \eqref{gcon} provides the conditions for higher-order decoupling. This condition is in general weaker than Eq. \eqref{totr}, so it enables more efficient sequence design. Additionally, higher-order robust state transfers can be achieved by minimizing components in the minimal subspace of $\bar H^{(r-1)}$ that does not commute with the initial density matrix.

Eq. \eqref{gcon} can also be used to achieve higher-order robustness to control imperfections including all cross terms by including the error Hamiltonians in $\lc{H_\text{pert}^w}$ \cite{chen2025engineering}. If the correlation time of the error Hamiltonians is finite, the errors are temporally localized. Consequently, imposing robustness against such errors does not alter the controllability of the system.

\section{Total cost function}\label{secfun}
Typically, the steps of designing a control sequence involves describing the system Hamiltonians, choosing a partitioning, specify control targets and use a search algorithm to find a sequence that meets the objectives by minimizing a total cost function \cite{chen2025engineering}:
\begin{equation}
	\begin{split}
		f_\text{tot}\lr{\lc{a_k(t)}}=&w_\text{pri}f_\text{pri}\lr{\lc{a_k(t)}}+w_0f^{(0)}\lr{\lc{a_k(t)}}\\
		&+\sum_{r=2}^Rw_{r-1}f^{(r-1)}\lr{\lc{a_k(t)}},
	\end{split}
	\label{totobj}
\end{equation}
where $w_\text{pri}$, $w_\text{0}$ and $w_{r-1}$ are positive weights reflecting relative importance of each objective. The cost functions are:
\begin{equation}
	\begin{split}
		&f_\text{pri}\lr{\lc{a_k(t)}}=1-\frac{|\text{Tr}\lr{U_\text{pri}(T_\text{seq})^\dagger U_\text{target}}|}{\text{Tr}\lr{U_\text{target}^\dagger U_\text{target}}},\\
		&f^{(0)}\lr{\lc{a_k(t)}}\\
		&=\left|	\int_0^{T_\text{seq}}dt_1|H_\text{tog}(t_1)\rrangle_\text{comp}-T_\text{seq}\bigoplus_{w=1}^W |s_wH_\text{target}^w\rrangle_w\right|,\\
		&f^{(r-1)}\lr{\lc{a_k(t)}}\\
		&=\ls{\sum_G|\mathbf A(G)\cdot \vec c^{(r-1)}-\vec d(G)|^2}^{1/2}.
	\end{split}
	\label{subobj}
\end{equation} 
The first two cost functions are the same as \cite{chen2025engineering}. The higher-order cost functions $f^{(r-1)}\lr{\lc{a_k(t)}}$ are obtained from Eq. \eqref{gcon}. This replaces the higher-order correction cost function introduced in \cite{chen2025engineering} that is based on Eq. \eqref{totr}. To setup the cost function one needs to specify the perturbative parameters $\lc{\eta_\text{pert}^m}$ and all inequivalent, nontrivial parameter graphs. For a desired target $\bar H_\text{target}^{(r-1)}$, its controllability can be determined by checking if $\partial_G\bar H_\text{target}^{(r-1)}\in\mathscr S(G)$ for all $G$. Then $f^{(r-1)}\lr{\lc{a_k(t)}}$ can be evaluated by calculating $\mathbf A(G)$ and $\vec d(G)$ in Eq. \eqref{gcon} for all $G$.
\section{Engineering $\bar H^{(r-1)}$ on qubit networks}\label{QN1}
	As an example application, we show how to engineer $\bar H^{(r-1)}$ on qubit networks, such as nearest-neighbor chains and square lattices. 
	These qubit networks appear in leading experimental platforms \cite{kjaergaard2020superconducting,haffner2008quantum,zeiher2016many,rechtsman2013photonic,blais2004cavity,kok2007linear}, and are canonical testbeds/models for quantum simulation \cite{georgescu2014quantum,houck2012chip,gross2017quantum,monroe2021programmable,blatt2012quantum,ebadi2021quantum}, Hamiltonian engineering/quantum control \cite{ajoy2012quantum,schulte2005optimal,khaneja2002sub,ajoy2019selective}, communication and state transfer \cite{bose2003quantum,bose2007quantum,burgarth2005conclusive,christandl2005perfect,wei2023quantum,godsil2012state}, quantum error correction and fault-tolerant computing \cite{fowler2012surface,kitaev2003fault,raussendorf2007fault,horsman2012surface,raussendorf2007topological}, and many-body dynamics \cite{imbrie2016many,schreiber2015observation,wei2018exploring,browaeys2020many,bloch2008many,lepoutre2019out,cassidy2011generalized,iyoda2018scrambling,deutsch1991quantum,bohigas1984characterization} and may be explored on NISQ processors \cite{preskill2018quantum,smith2019simulating,cervera2018exact,chowdhury2024enhancing,alam2025programmable}.

\begin{figure}
		\centering
		\includegraphics[width=0.45\textwidth]{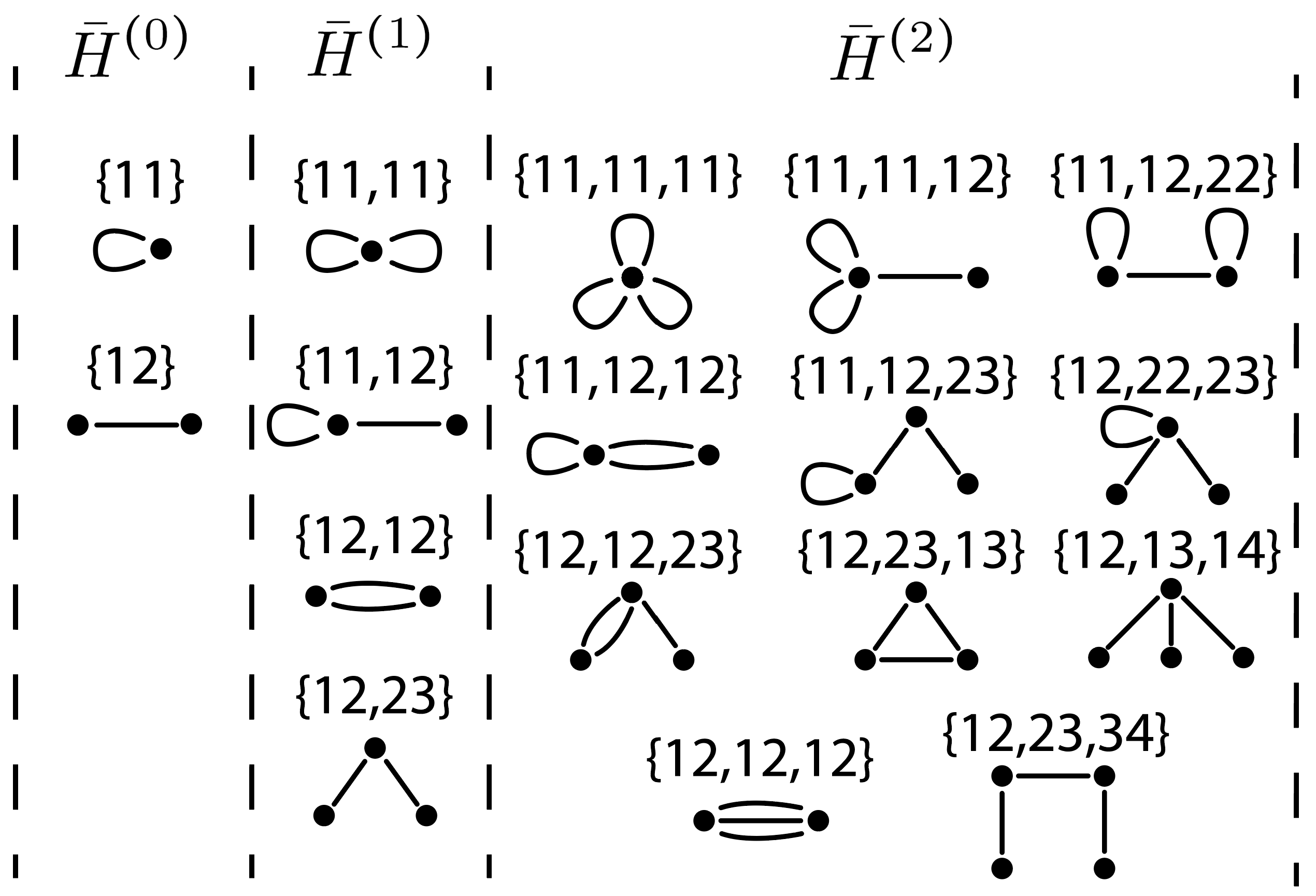}\\
		\caption{Nonisomorphic parameter graphs associated with Magnus terms.}
		\label{graphs}
	\end{figure}

\begin{figure}
		\centering
		\includegraphics[width=0.35\textwidth]{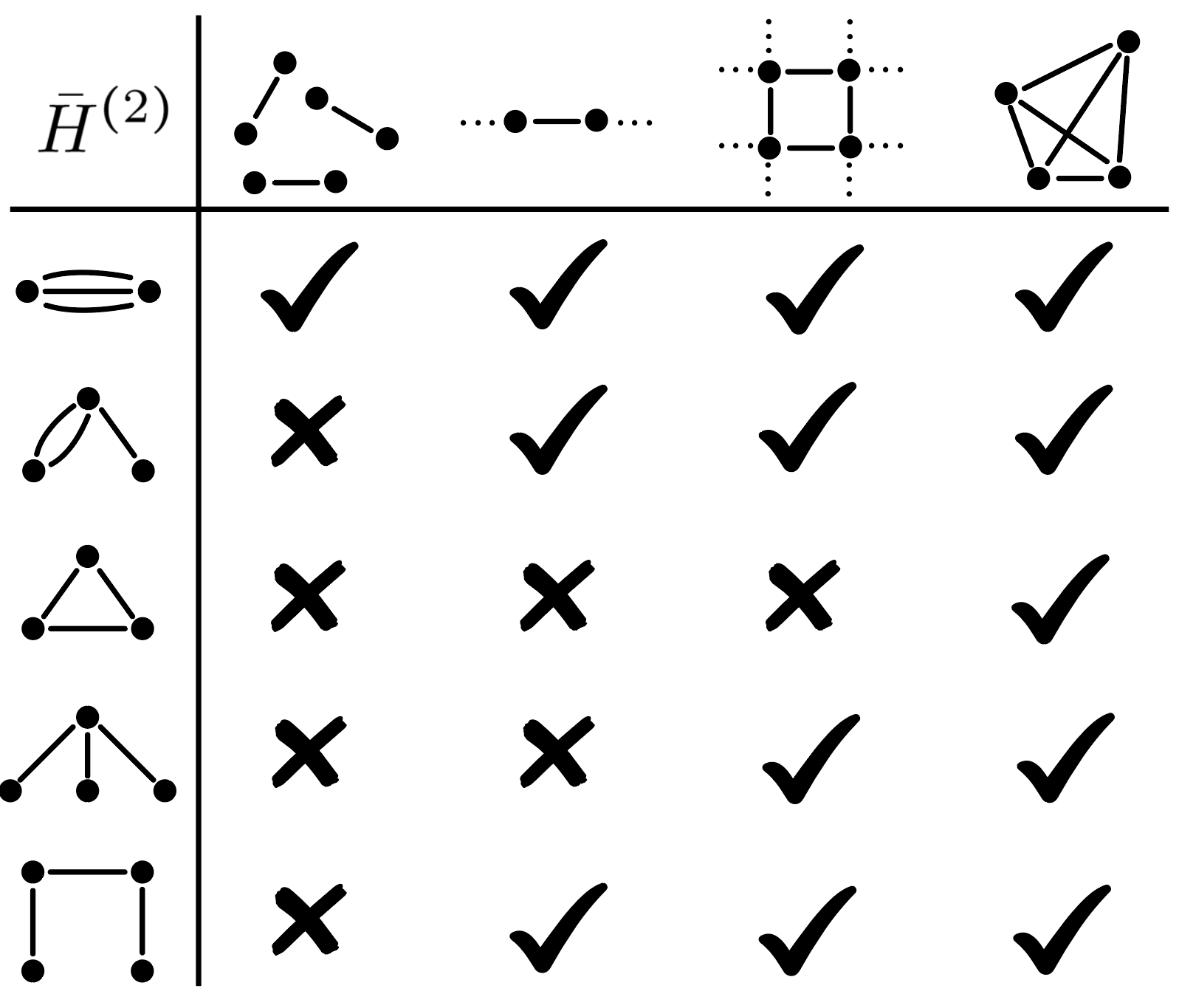}\\
		\caption{Networks with different topology and the parameter graphs that need to be considered for $\bar H^{(2)}$. From left to right: qubit pairs, nearest-neighbor chain, square lattice, and all-to-all ensemble. Only the parameter graphs with no loops are shown here.}
		\label{topology}
	\end{figure}

	We consider collective control of the qubits of a very common form
	\begin{equation}
		\begin{split}
			H_c(t)=&\frac{\omega_1(t)}{2}\ls{\cos(\phi(t))\sum_{i=1}^N\x^i+\sin(\phi(t))\sum_{i=1}^N\y^i}\\
			&+\frac{\Delta\omega(t)}{2}\sum_{i=1}^N\z^i,
		\end{split}
		\label{conH}
	\end{equation}
	where $\omega_1(t)$, $\phi(t)$ and $\Delta\omega(t)$ are the time-dependent amplitude, phase and detuning of the control field in a common reference frame (the rotating frame). This is a special case of the control Hamiltonian in Eq. \eqref{totH} with one control channel and identity transfer function. $N$ is the number of qubits in the network.
	The internal Hamiltonian is
	\begin{equation}
		\begin{split}
			H_\text{int}&=\frac{1}{2}\sum_{i=1}^N\delta_i\z^i+\frac{1}{4}\sum_{i<j}^NB_{ij}\vec\sigma_i\cdot\mathbf{D}\cdot\vec\sigma_j\\
		\end{split}
	\label{hint}
	\end{equation}
	where $\vec\sigma_i=(\x^i,\y^i,\z^i)$ and 
	\begin{equation}
		\sigma_\alpha^i=\mathbb 1^{\otimes(i-1)}\otimes\sigma_\alpha\otimes\mathbb 1^{\otimes(N-i)},
	\end{equation}
	for $\alpha=\lc{x,y,z}$,
	are the Pauli operators acting on the $i$th qubit. $\lc{B_{ij}}$ are coupling strengths, $\mathbf{D}$ is a rank-2 tensor determining the form of the qubit interactions, and $\delta_i$ is the strength of the qubit-dependent energy offset. $\lc{B_{ij}}$ and $\delta_i$ are random variables.
	The primary and perturbative Hamiltonians are defined as $H_\text{pri}(t)=H_c(t)$ and $H_\text{pert}=H_\text{pert}^1+H_\text{pert}^2$ with
	\begin{equation}
		H_\text{pert}^1=\frac{1}{2}\sum_{i=1}^N\delta_i\z^i,~~H_\text{pert}^2=\frac{1}{4}\sum_{i<j}^NB_{ij}\vec\sigma_i\cdot\mathbf{D}\cdot\vec\sigma_j.
	\end{equation} 
	The perturbative parameters are
	\begin{equation}
		\eta_\text{pert}^{ij}\equiv\begin{cases}
			B_{ij}&i\ne j\\
			\delta_i&i=j
		\end{cases}.
	\label{perpa}
	\end{equation}
	The network can be represented by a graph $G_\text{tot} = (V(G_\text{tot}), E(G_\text{tot}))$, where the vertex set $V(G_\text{tot})$ corresponds to the qubits and the edge set $E(G_\text{tot})$ to their connectivity/coupling. Then the parameter graphs are weakly isomorphic to subgraphs of $G_\text{tot}$ \cite{Note2}. The following statements hold for a parameter graph $G$:
	\begin{enumerate}
		\item $G$ is not connected implies $\partial_G \bar H^{(r-1)}=0$;
		\item $\partial_G \bar H^{(r-1)}\ne0$ implies $G$ has $r$ edges;
		\item Two parameter graphs are equivalent iff they are isomorphic.
	\end{enumerate}
	Therefore, for each $\bar H^{(r-1)}$, only nonisomorphic, connected parameter graphs with $r$ edges are considered.
	Fig. \ref{graphs} shows nonisomorphic graphs for different $\bar H^{(r-1)}$ for a network with all-to-all interactions. Fig. \ref{topology} shows nonisomorphic graphs for $\bar H^{(2)}$ for qubit systems with different topology.

The evaluation of $\mathscr S(G)$ and $\mathbf A(G)$ requires calculating $\partial_G F_{i_1\cdots i_r}$, allowing characterizing controllabilty of $\bar H^{(r_1)}$ and defining the objective functions $f^{(r-1)}$. $\partial_G F_{i_1\cdots i_r}$ can be calculated, for example, as
\begin{equation}
	\begin{split}
	\partial_{\lc{11,12}}F_{i_1i_2}&=\frac{\partial^2 F_{i_1i_2}}{\partial\delta_1\partial B_{12}}\\
	&=F\lr{\frac{\partial h_{i_1}}{\partial \delta_1},\frac{\partial h_{i_2}}{\partial B_{12}}}+F\lr{\frac{\partial h_{i_1}}{\partial B_{12}},\frac{\partial h_{i_2}}{\partial \delta_1}},
	\end{split}
\end{equation}
where $F(h_{i_1},\ldots,h_{i_r})$ is defined by Eq. \eqref{Fle} with $h_{i_1},\ldots,h_{i_r}$ being variables.  

In the presence of systematic control error described by an error Hamiltonian
\begin{equation}
	\Delta H=\sum_i\epsilon_i H_{i}(t),
\end{equation}
where $\epsilon_i$ are random variables and $H_{i}(t)$ are time-dependent variations of $H_c(t)$ that act collectively on the qubits, the perturbative Hamiltonian can include $\lc{\epsilon_i H_{i}(t)}$ and the additional perturbative parameters are $\eta_\text{pert}^{e_i}\equiv\epsilon_i$. In this case, the parameter graphs can include $\lc{e_i}$ as ``edges". In addition to the ones listed in Fig. \ref{graphs}, the inequivalent parameter graphs include
\begin{equation}
\begin{split}
	\bar H^{(0)}:&~\lc{e_i},\\
	\bar H^{(1)}:&~\lc{e_i,11},\lc{e_i,12},\lc{e_i,e_j},\\
	\cdots.&
\end{split}
\end{equation}
\section{Examples}\label{secexa}
In this section, we show how to use the method for higher-order engineering to achieve advantage in decoupling, simulation and spectroscopy. We focus on networks with all-to-all interactions, for which all inequivalent parameter graphs must be considered; the solutions obtained therefore apply to more restricted topologies as well. For specific network structures, more efficient designs may be achieved by excluding parameter graphs that are irrelevant to the given topology.

For control imperfection, consider Rabi strength variation
\begin{equation}
	\Delta H(t)=(1+\epsilon)H_c(t),
\end{equation}
where $\epsilon$ is a dimensionless random variable, typically between -1 and 1. This imperfection is one of the most ubiquitous errors in quantum control and has been widely considered in robust control and decoupling sequences. Consider the internal Hamiltonian in Eq. \eqref{hint} with
\begin{equation}
	\mathbf D=\text{Diag}(-1,-1,2).
\end{equation}
So, the interactions correspond to dipolar interactions under high field which is the most studied interaction type in AHT. The choice of Rabi strength variation and dipolar interactions makes it easier to compare our results with other well-known, state-of-the-art sequences. But of course, the method presented here can be applied to other error and interaction types \cite{chen2025engineering}.

The primary and perturbative Hamiltonians are $H_\text{pri}(t)=H_c(t)$ with $H_c(t)$ given by Eq. \eqref{conH} and $H_\text{pert}=H_\text{pert}^1+H_\text{pert}^2+H_\text{pert}^3$ where
\begin{equation}
	\begin{split}
	&H_\text{pert}^1=\epsilon H_c(t),\\
	&H_\text{pert}^2=\frac{1}{2}\sum_{i=1}^N\delta_i\z^i,\\
	&H_\text{pert}^3=\frac{1}{4}\sum_{i<j}^NB_{ij}\vec\sigma_i\cdot\mathbf{D}\cdot\vec\sigma_j.
	\end{split}
\end{equation}
So the perturbative parameters are $\eta_\text{pert}^e\equiv\epsilon$ and those defined by Eq. \eqref{perpa}. The $\mathscr C$ spaces can be calculated using the method from \cite{chen2025engineering}, their dimensions are 3, 3 and 5. So the composite $\mathscr C$ space is 11-dimensional. Because our goal is to highlight coherent Hamiltonian engineering, we neglect relaxation in these examples, even though it ultimately limits the achievable performance (e.g., via the relaxation time in the rotating frame $T_{1\rho}$~\cite{abragam1961principles,yan2013rotating}).

Unless stated otherwise, each sequence is piecewise-constant with segment duration of $\Delta t = 2~\mu\text{s}$. A sequence with $Q$ segments therefore has a total duration $T_\text{seq}= Q\Delta t$. 
We consider $0<\omega_1(t)<\omega_\text{max}$, $-\omega_\text{max}<\Delta\omega(t)<\omega_\text{max}$ with $\omega_\text{max}=250$ kHz. $\omega_1(t)$ and $\Delta\omega(t)$ of the first and last segments of each sequences are set to zero to allow easy composition of sequences \cite{fortunato2002design}. Although, we consider the more general case where $\Delta\omega(t)$ is time-dependent, all the sequences here can also be designed with on-resonance segments ($\Delta\omega(t)\equiv0$) which is somewhat more familiar to the quantum optimal control community. Generally, on-resonance sequences are easier to design due to a smaller size of the solution space but less efficient than their off-resonance counterparts. We also choose $U_\text{target}=\mathbb{1}$ and $H_\text{target}^w=0$ (except for the qubit storage sequence) for all examples so they only differ in the choice of $f^{(r-1)}$ ($r>1$).

We use CMA-ES \cite{hansen2016cma} because it performs well on nonconvex, multimodal objective landscapes and is efficient in terms of function evaluations. However, the objective functions are not tied to this choice and can be optimized with any general-purpose optimizer. In practice, we begin the optimization with a small $Q$ and progressively increase its value until a satisfactory solution is found, which accounts for the varying sequence lengths across the examples.

\subsection{Decoupling}\label{ex1}
The first example is a decoupling sequence that removes zeroth- and first-order average Hamiltonians. It also removes second-order average Hamiltonians of the dipolar interaction since this term is typically stronger than Rabi field variations and qubit detunings. Of course, including higher-order robustness to detunings is relatively easy and can be done with composite pulse techniques.

\begin{figure}
	\centering
	\includegraphics[width=0.5\textwidth]{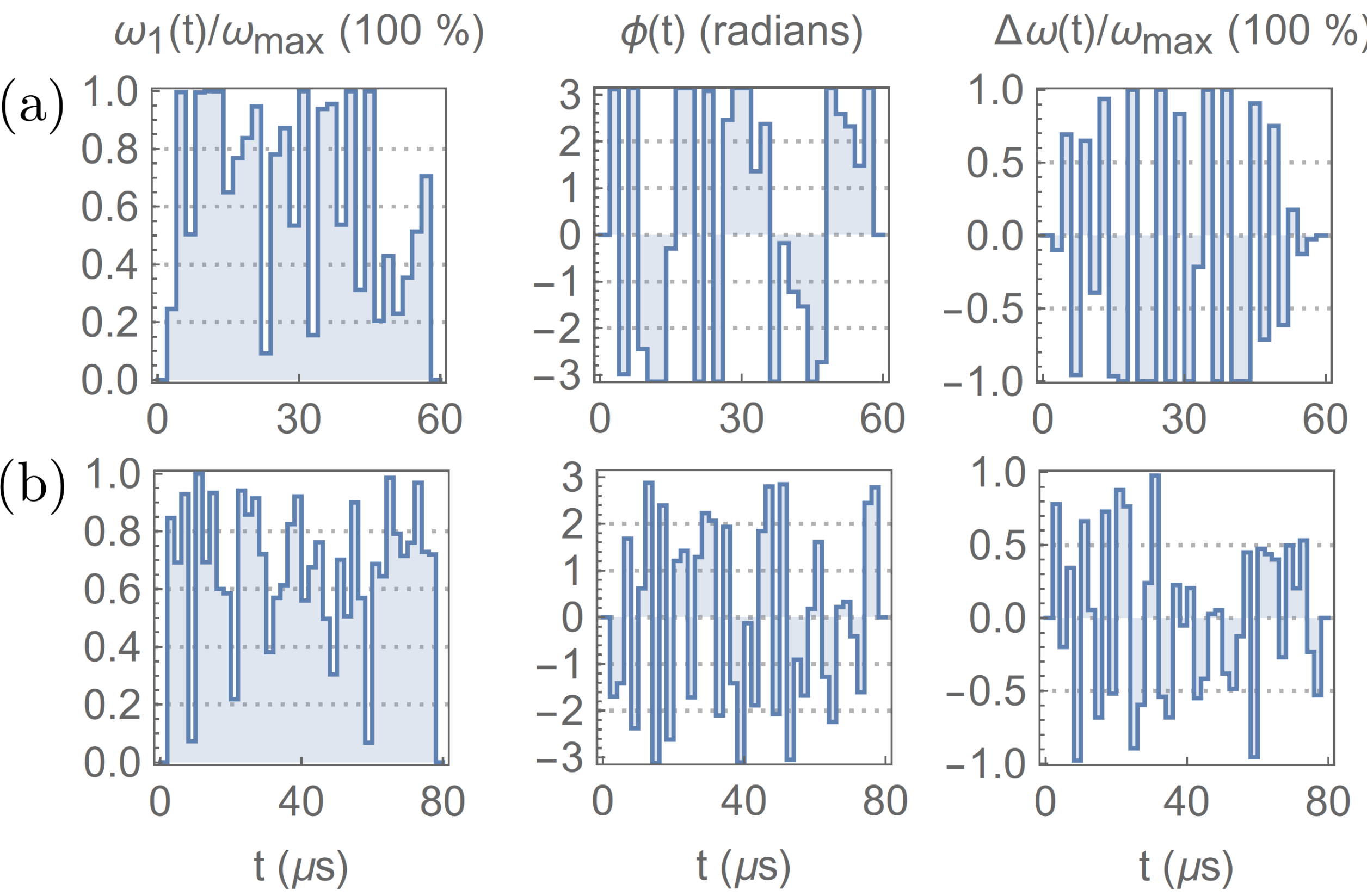}\\
	\caption{Control parameters of the sequences (a) $P_\rho$ and (b) $P_\text{uni}$.}
	\label{pplot1}
\end{figure}

The inequivalent parameter graphs for $\bar H^{(1)}$ are
\begin{equation}
	\begin{split}
	&\lc{11,11},\lc{11,12},\lc{12,12},\lc{12,23},\\
	&\lc{e,11},\lc{e,12},\lc{e,e}.
	\end{split}
\end{equation}
The ones for $\bar H^{(2)}$ for dipolar interactions are
\begin{equation}
	\begin{split}
	&\lc{12,12,23},\lc{12,23,13},\lc{12,13,14},\\
	&\lc{12,12,12},\lc{12,23,34}.
	\end{split}
\end{equation}
Since the goal is decoupling, the targets are $\vec d(G)=0$ for all parameter graphs for both first and second orders.
The sequence ($P_\text{uni}$) is shown in Fig. \ref{pplot1} (b). As a comparison, we also design a qubit storage sequence, $P_\rho$, that only removes the components at each order that do not commute with $\rho_0$. This results in a more efficient sequence (Fig. \ref{pplot1} (b)). The sequences can be further symmetrized ($P_\text{uni}^\text{sym}$ and $P_\rho^\text{sym}$) to remove the third-order corrections that do not involve $\epsilon$.

\begin{figure}
	\centering
	\includegraphics[width=0.45\textwidth]{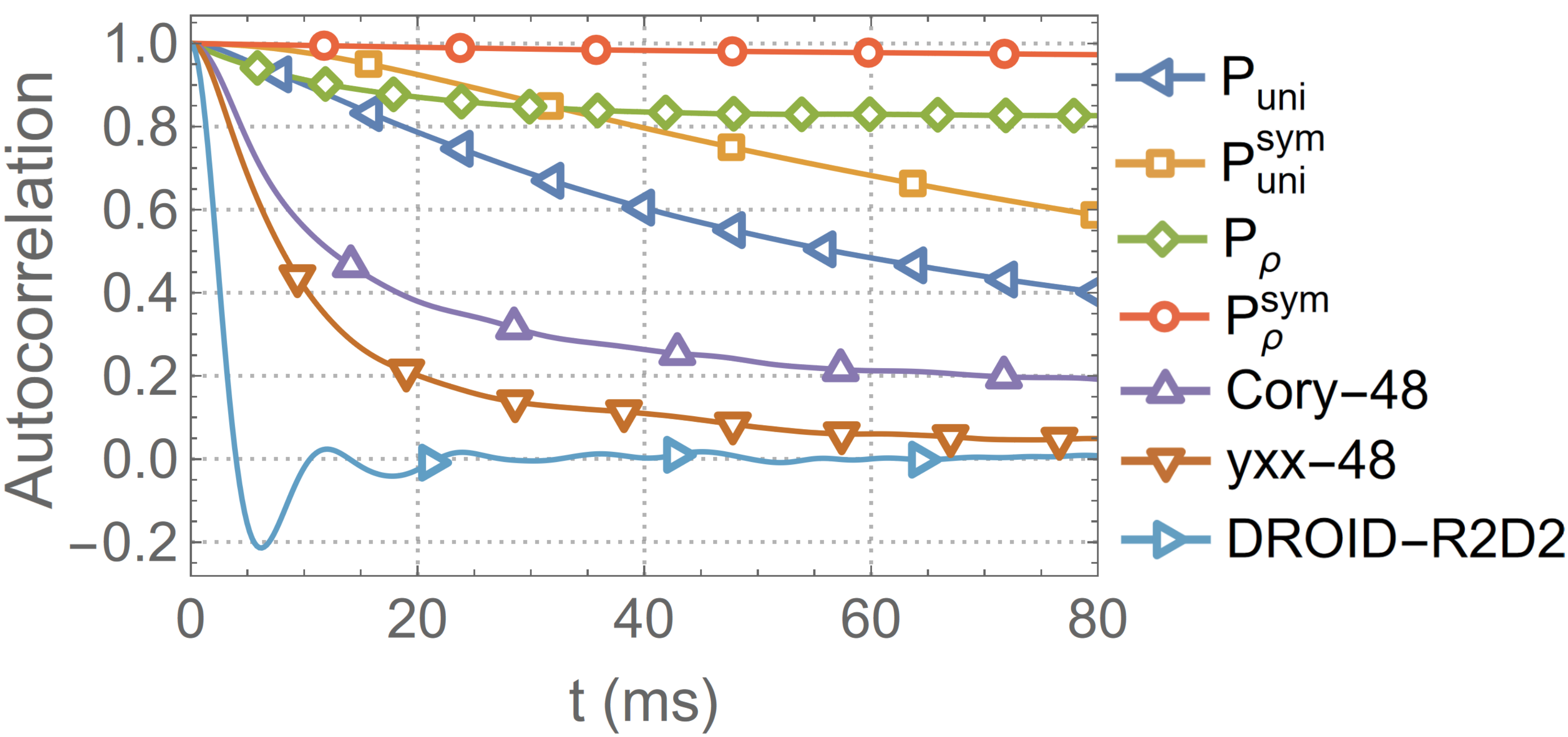}\\
	\caption{Autocorrelation of $\rho_0=\sum_{i=1}^N\x^i$ under different sequences. The results are obtained from simulations with a 6-qubit ensemble, averaged over 100 realizations.}
	\label{zhorg}
\end{figure}

To test the performance of the sequence, we simulate the autocorrelation function of the initial state $\rho_0=(\sum_{i=1}^N\x^i)$. The identity part of the density matrix has been omitted since it does not contribute to the signal. This initial state corresponds to an ensemble polarized along $x$, which is commonly used to test decoupling sequences. The autocorrelation corresponds to the real part of the signal obtained from stroboscopic quadrature detection
\begin{equation}
	S(t=kT_\text{seq})=\frac{\text{Tr}(U(T_\text{seq})^k\rho_0U(T_\text{seq})^{\dagger k}\rho_0)}{N2^N},
\end{equation}
where $(U(T_\text{seq})$ is calculated according to Eq. \eqref{ute} and $k$ is the number of sequences applied.
We compare the results with alternative state-of-the-art dipolar decoupling sequences \cite{cory1990time,peng2022deep,choi2020robust,tyler2023higher}. All the sequences operates at the maximal Rabi frequency $\omega_\text{max}$ (1 $\mu s$ for a $\pi/2$ rotation). We set $\tau=2~\mu\mathrm{s}$ as the in-sequence delay: it is the delay in each solid-echo subcycle of Cory-48 and the window duration in yxx-48 and DROID-R2D2. The perturbative parameters $B_\text{ij}$, $\delta_i$ and $\epsilon$ are drawn from independent zero-mean normal distributions of standard deviations $\sigma_\text{dip}$, $\sigma_\text{z}$ and $\sigma_\epsilon$ respectively. Here, we consider the ``moderate error" regime: $\sigma_\text{dip}=0.02\omega_\text{max}=5$ kHz, $\sigma_z=0.01\omega_\text{max}=2.5$ kHz and $\sigma_\epsilon=0.02$. The results are shown in Fig. \ref{zhorg}. 
The sequences designed using our method greatly outperform the other sequences, demonstrating their superb performance in coherent averaging. In particular, the coherence time achieved by $P_\rho^\text{sym}$ is more than $10^2$ times longer than that of Cory-48. Performance comparison in other regimes is in Appendix \ref{appen3}.

\subsection{Engineering 3-body Hamiltonian}\label{exam2}
This example designs a robust sequence that achieves a three-body interaction at first order.
Higher-order corrections are less important when engineering nonzero target effective Hamiltonian \cite{chen2025engineering}. So, in this case, we do not consider second-order corrections. 

All parameter graphs that involve detuning or Rabi strength variation need to be decoupled:
\begin{equation}
	\lc{11,11},\lc{11,12},\lc{e,11},\lc{e,12},\lc{e,e}.
\end{equation}
We also choose to decouple $\lc{12,12}$ since the subspace of this graph only includes single-body terms. So, the minimal subspace of $\lc{12,23}$ is calculated with $\mathscr C$-integrals restricted in the nullspace of $\mathbf A(\lc{12,12})$, and it is spanned by
\begin{equation}
	\begin{split}
		\{
		&xyz+xzy+yxz+yzx+zxy+zyx,\\
		&xxy+xyx+yxx-2yyy+zzy+zyz+yzz,\\
		&xxz+xzx+zxx-2zzz+yyz+yzy+zyy,\\
		&yyx+yxy+xyy-3xxx+2zzx+2zxz+2xzz,\\
		&-xxy-xyx-yxx+zzy+zyz+yzz,\\
		&xxz+xzx+zxx-yyz-yzy-zyy,\\
		&-3yyx-3yxy-3xyy+xxx+2zzx+2zxz+2xzz\}.
	\end{split}
\end{equation}
Here, we use the shorthands $x,y$ and $z$ to represent the Pauli matrices. Normalization constants and the tensor product symbols have been omitted. The above Hamiltonians only represent the ``form" of the actual effective Hamiltonians. They can be understood as the Hamiltonians of the first three qubits from the network.

\begin{figure}
	\centering
	\includegraphics[width=0.5\textwidth]{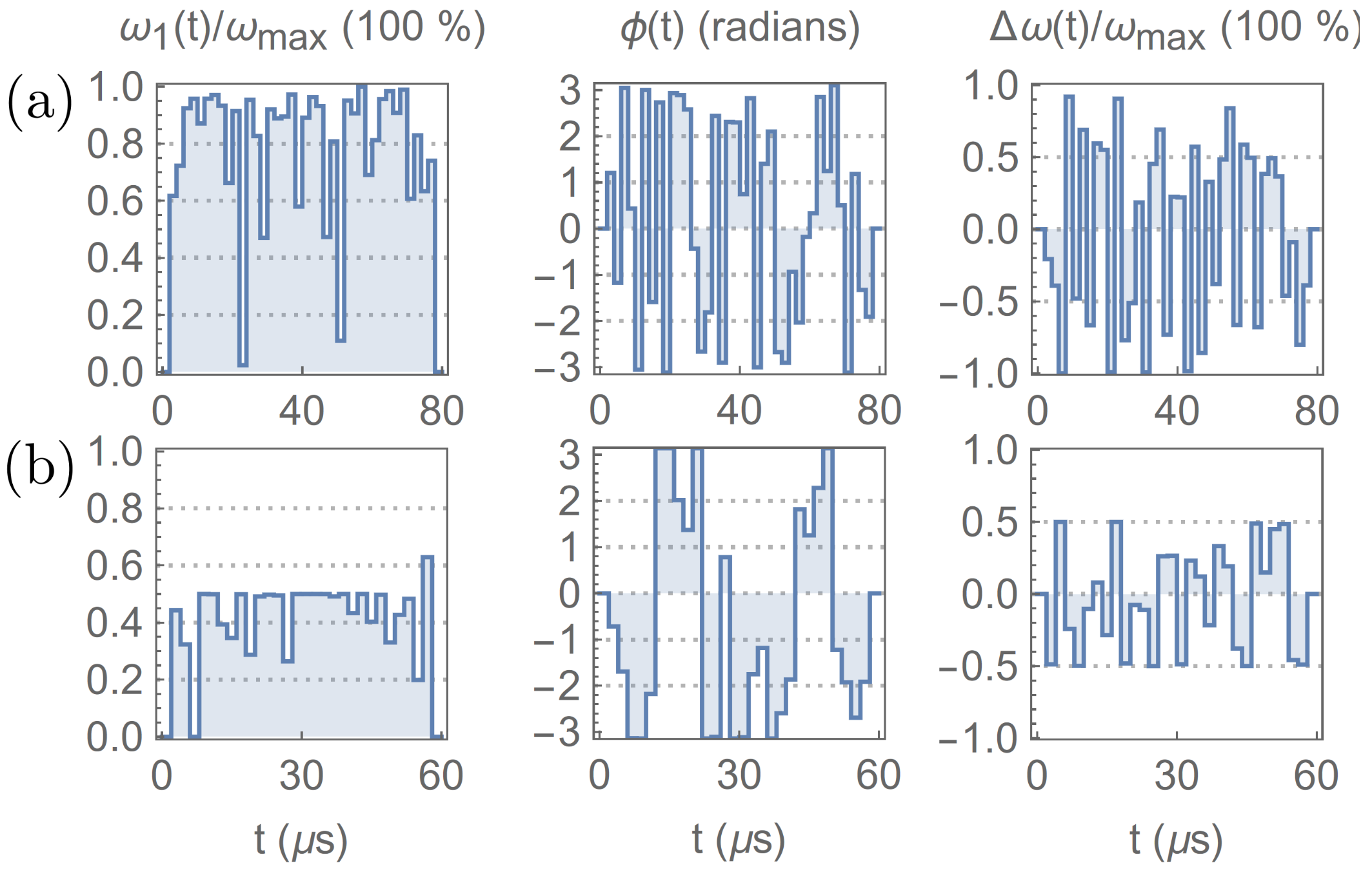}\\
	\caption{Control parameters of the sequences (a) $P_{xyz}$ and (b) $P_\text{cross}$.}
	\label{pplot2}
\end{figure}

As an example, we choose to engineer the first term. No constraint is put on the sign of the dynamics. The sequence ($P_{xyz}$) is shown in Fig. \ref{pplot2} (a). 
It achieves the effective dynamics
\begin{equation}
	\begin{split}
		&-T'\sum_{i<j<k}(B_\text{ij}B_\text{jk}+B_\text{ij}B_\text{ik}+B_\text{ik}B_\text{jk})\\
		&\big(\x^i\y^j\z^k+\x^i\z^j\y^k+\y^i\x^j\z^k\\
		&~~~~~~~~~+\y^i\z^j\x^k+\z^i\x^j\y^k+\z^i\y^j\x^k\big)
	\end{split}
\end{equation}
with $T'=18.4$ (recall that the units are absorbed in the $\mathscr C-$integrals).

\begin{figure}
	\centering
	\includegraphics[width=0.3\textwidth]{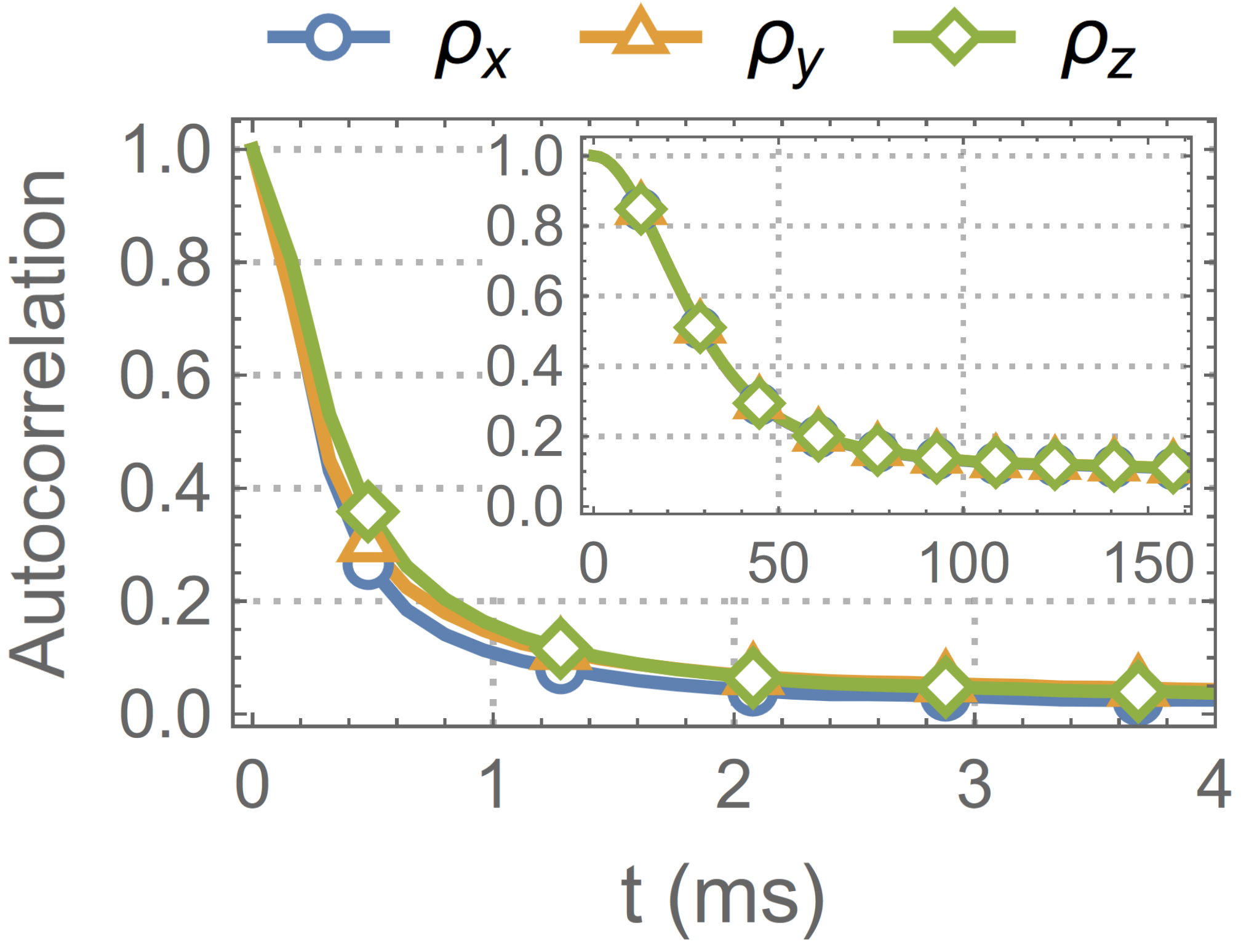}\\
	\caption{Autocorrelation of the states $\rho_x$, $\rho_y$ and $\rho_z$ under the sequence $P_\text{xyz}$. The inset corresponds to $\sigma_\text{dip}=\sigma_z=0.01\omega_\text{max}$. The main figure  corresponds to $\sigma_\text{dip}=\sigma_z=0.1\omega_\text{max}$. The results are simulated with a 6-qubit ensemble, averaged over 500 realizations. }
	\label{3bud}
\end{figure}

To test its performance, we simulate the autocorrelation of $\rho_\alpha=\sum_{i=1}^N\sigma_\alpha^i/2$, for $\alpha=x,y$ and $z$. The states have the same decay profile under the effective Hamiltonian due to its symmetry with respect to permutation of $x,y$ and $z$. Fix $\sigma_\epsilon = 0.01$. We compare two cases: (1) $\sigma_{\mathrm{dip}}=\sigma_z=0.01\,\omega_{\max}$ and (2) $\sigma_{\mathrm{dip}}=\sigma_z=0.1\,\omega_{\max}$.
The results are shown in Fig. \ref{3bud}. 
In case (1), the decay curves of $\rho_x$, $\rho_y$ and $\rho_z$ overlap nearly perfectly, confirming the precision of the engineered effective Hamiltonian. Discrepancies appear in case (2), where the larger parameter dispersion increases higher-order corrections and may push the system beyond the perturbative regime.

\subsection{Correlation spectroscopy through first-order effective Hamiltonians}\label{exam3}

Chemical shift anisotropy (CSA)/dipolar cross terms have been analyzed in several contexts, including second-order recoupling under MAS \cite{ernst1996second}, coherent symmetry-breaking/lineshape distortions in solids \cite{duma2003carbon}, and cross-correlated relaxation (relaxation interference) in solution NMR \cite{tessari1997quantitative}. This example designs a sequence that systematically isolates cross-term contributions while suppressing other undesired terms.

We need to decouple
\begin{equation}
		\lc{11,11},\lc{12,12},\lc{12,23},\lc{e,11},\lc{e,12},\lc{e,e},
\end{equation}
while engineering a target in $\lc{11,12}$. The space is spanned by
\begin{equation}
	\lc{2zz-xx-yy,xx-yy,xy,yx,xz,zx,yz,zy}
	\label{crossbasis}
\end{equation}
Note, $xy$ is different from $yx$, since
\begin{equation}
	\begin{split}
		xy\rightarrow\sum_{i<j}B_{ij}\lr{\delta_i\x^i\y^j+\delta_j\y^i\x^j},\\
		yx\rightarrow\sum_{i<j}B_{ij}\lr{\delta_i\y^i\x^j+\delta_j\x^i\y^j}.
	\end{split}
\end{equation}
We choose to engineer the first term, $2zz-xx-yy$, from \eqref{crossbasis}.
This target has effective coupling strength that depend on both the original coupling strength and detunings of qubits incident to the coupling. Therefore, correlation between $B_{ij}$ and $\delta_i+\delta_j$ can be detected by measuring evolution of the system under this effective Hamiltonian.

We design a sequence ($P_\text{cross}$, Fig. \ref{pplot2} (b)) of 30 segments that achieves a total action
\begin{equation}
	-T'\sum_{i<j}B_{ij}(\delta_i+\delta_j)(2\z^i\z^j-\x^i\x^j-\y^i\y^j)
\end{equation}
with $T'=16\sqrt{12}$.

\begin{figure}
	\centering
	\includegraphics[width=0.5\textwidth]{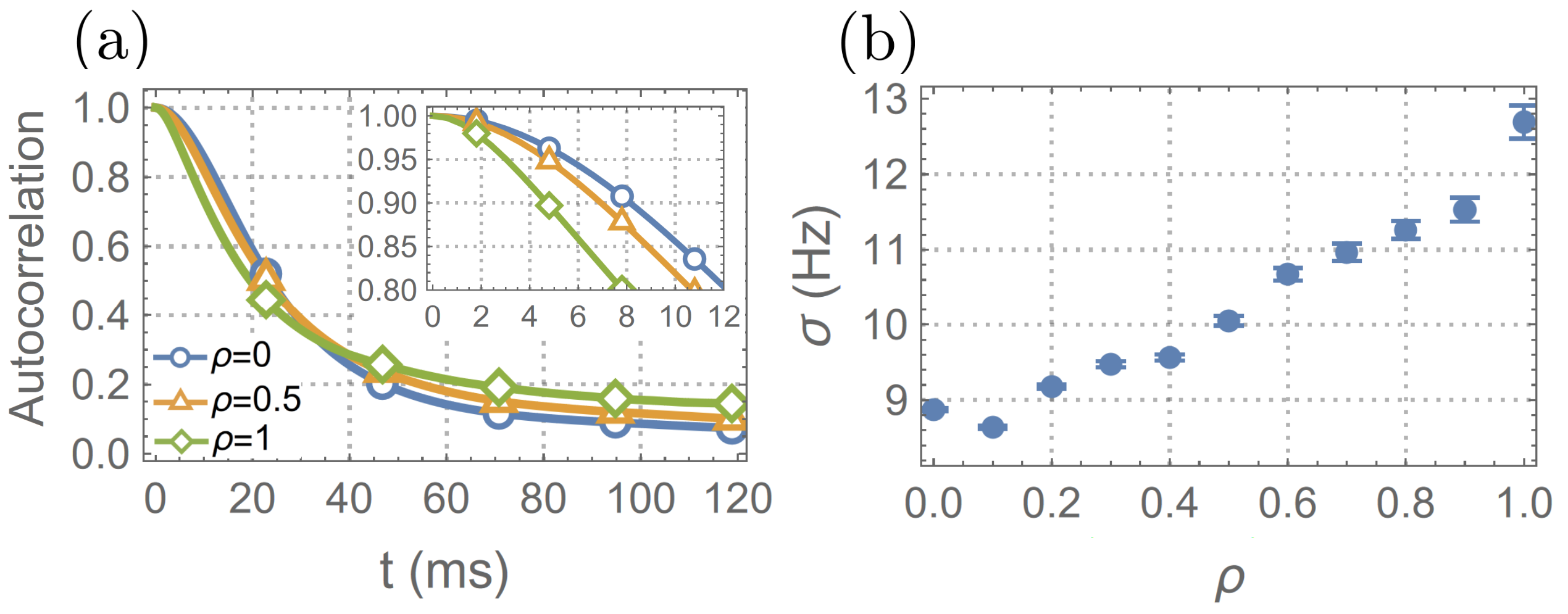}\\
	\caption{(a) Autocorrelation of $\rho_0=\sum_{i=1}^N\x^i$ under the sequence for engineering CSA/dipolar cross term for $\rho=0,~0.5$ and 1. The results are obtained from simulations of a 6-qubit ensemble, averaged over 500 realizations.
	(b) Standard deviation of the initial Gaussian decay by fitting the early-time signal (up to $12$ ms) to $\exp(-t^2\sigma^2/2)$.
	}
	\label{crossy}
\end{figure}

To test its performance, we simulate the evolution of $\rho_0=\sum_{i=1}^N\x^i$ on a 6-qubit network. Choose $\sigma_z=\sigma_\text{dip}=0.01\omega_\text{max}$ and $\sigma_\epsilon=0.01$, and let
\begin{equation}
	\begin{split}
		&\epsilon\sim\mathcal N(0,\sigma_\epsilon^2),~~\delta_i\sim\mathcal N(0,\sigma_z^2),\\
		&B_{ij}\sim\rho\frac{\sigma_\text{dip}}{\sqrt2\sigma_z}(\delta_i+\delta_j)+\mathcal N(0,(1-\rho^2)\sigma_\text{dip}^2).
	\end{split}
\end{equation}
The marginal distribution of $B_{ij}$ is $\mathcal N(0,\sigma_\text{dip}^2)$, and $\rho$ is the correlation between $B_\text{ij}$ and $\delta_i+\delta_j$. The results are shown in Fig. \ref{crossy}. As the correlation $\rho$ increases, the average effective coupling strength $B_{ij}(\delta_i+\delta_j)$ becomes larger, resulting in spectra with faster initial decay. The tails of the decays, however, become longer for increased $\rho$, indicating a transition from Gaussian decays to exponential decays.

\section{Discussion and outlook}\label{secdis}
The simple examples above illustrate capabilities that were previously difficult to achieve in a systematic way: (1) the design of highly robust decoupling sequences; (2) the engineering of arbitrary, robust higher-order effective Hamiltonians (e.g., three-body interactions); and (3) the engineering of controlled nonlinear dependence on internal parameters. Collectively, these examples highlight how the method provides advantages in decoupling, simulation and spectroscopy. Although the examples use shaped pulses with continuous control parameters, the objective functions in Eq. \eqref{subobj} can evaluate the performance of any control sequence. The same framework therefore applies to sequences constructed from a finite gate set (e.g., collective $\pi/2$ rotations), where the optimization can be carried out with a combinatorial optimizer \cite{chen2025control}.

In Example \ref{ex1}, the sequences $P_\text{uni}$, $P_\text{uni}^\text{sym}$, $P_\rho$ and $P_\rho^\text{sym}$ greatly outperform state-of-the-art decoupling sequences. This improvement stems from the objective function $f^{(r-1)}$, which exactly captures the contributions of higher-order terms, including cross terms, at each order, without introducing additional approximations. In the example, the vectorized 
$\mathscr C$-integrals are 11-dimensional at zeroth order and 121-dimensional at first order; at second order, even restricting to dipolar interactions, the space is 125-dimensional. Manually crafting a sequence that simultaneously controls all such terms is therefore infeasible. In contrast, brute-force numerical optimization of coherence time can be highly inefficient, as it provides little structural guidance for the search.

Interestingly, the state storage sequences exhibit ($P_\rho$, $P_\rho^\text{sym}$) distinctive features from the universal decoupling sequences ($P_\text{uni}$, $P_\text{uni}^\text{sym}$). Namely, they generally give longer coherence time and have exponential initial decay rather than Gaussian. This can be due to a second-averaging effect of the subspace orthogonal to the initial state. By not suppressing error terms that commute with the initial state (or, more generally, with the projector onto the information-bearing subspace), the orthogonal complement evolves under its own effective internal dynamics, decoupling it from the initial subspace and removing correlations between the two. This leads to a longer coherence time and a Gaussian-exponential decay transition \cite{kubo1963stochastic}.

In practice, the performance of the sequences is limited by decoherence and relaxation of the system. Markovian noise can be partially accounted for using the average Liouvillian formula \cite{ghose1999average} while effects of correlated random noise can be reduced using the cumulant expansion \cite{kubo1962generalized,chen2025engineering}. But generally, decoherence or relaxation due to random fluctuations cannot be completely suppressed. It will also be important to account for other types of control imperfections (e.g., transient distortions and nonlinearities) when relevant.

Unlike decoupling, there are far fewer established examples that deliberately target nonzero higher-order effects, and we are not aware of widely used benchmark sequences for direct comparison. In fact, without controllability characterization, it is highly nontrivial what effective Hamiltonians can be achieved at all at higher orders. 
The simple demonstrations presented here merely suggest potential uses in simulation and spectroscopy; identifying additional applications enabled by this framework is an interesting direction for future work.

As in the decoupling setting, our sequences for generating a three-body interaction and for engineering a detuning/interaction cross term control the full set of accompanying cross terms, and can incorporate higher-order corrections (e.g., second-order terms, as demonstrated in the decoupling example) when desired. This robustness is generally crucial because higher-order effective Hamiltonians are intrinsically small; even modest residual errors can dominate the intended effect.

Like any perturbative approach, the method can fail when the errors exceed the perturbative regime. Therefore, it is important to develop a framework that operates beyond the perturbative regime.
Brute-force optimization, albeit with poor efficiency, may uncover sequences that remain effective beyond the perturbative regime.
As shown in Sec.~\ref{QN1}, to design a sequence that is correct through order $r-1$, it suffices to analyze an $(r+1)$-qubit network. Consequently, simulations or experiments on $(r+1)$-qubit networks are, in principle, sufficient to validate an $(r-1)$th-order control sequence. The set of sequences that perform well uniformly over all $(r+1)$-qubit networks is generally larger than the set obtained by explicitly targeting the Magnus expansion through order $r-1$, yet smaller than the set that performs well uniformly over larger networks (i.e., $(r'+1)$ qubits with $r'>r$). Exploring this hierarchy may clarify how control sequences transition from perturbative designs to genuinely nonperturbative solutions.

We did not attempt to maximize the total action in Examples~\ref{exam2} and \ref{exam3}. In practice, maximizing the total action can be important: it improves efficiency and typically enhances robustness to experimental imperfections by increasing the separation between the desired effect and residual errors. Unlike the zeroth-order average Hamiltonian \cite{chen2025engineering}, the achievable range of higher-order actions cannot, in general, be characterized by a simple linear program. Developing efficient methods to quantify and optimize these higher-order action bounds may therefore be crucial. 

Although the above examples engineer the Magnus expansion term by term, some achievable effective Hamiltonians arise only as combinations of contributions from multiple orders. Such targets can be handled within the same approach by selecting objective functions $f^{(r-1)}$ that encode the desired combined outcome. However, imposing nonzero targets at different orders can introduce trade-offs and mutual constraints, which may reduce the set of achievable solutions. A general method for characterizing controllability under simultaneous multi-order targets is therefore needed.
	\section{Conclusion}\label{secon}
	This work extends the framework introduced in \cite{chen2025engineering} to enable engineering robust and precise higher-order effects in general Hamiltonian control systems. A central application is Hamiltonian engineering on qubit networks, which we illustrate through examples in decoupling, simulation, and spectroscopy. The decoupling example significantly outperforms widely used state-of-the-art sequences, while the engineered three-body dynamics and detuning/interaction cross terms highlight the broader scope of the approach beyond decoupling.
	
	The method identifies the exact achievable Hamiltonian/error space at each order, enabling targeted optimization of any desired achievable effective dynamics. Parameter graphs, closely tied to the topology of the qubit network, specify which parameters are retained symbolically while the remaining parameters enter only quantitatively. From these graphs, both the space of achievable targets and the corresponding objective functions for Hamiltonian engineering can be derived systematically. This general, intuition-independent procedure enables the design of practical, high-performance control sequences tailored to system-specific requirements and targets.
	
	This work extends the practical utility of average Hamiltonian theory and represents a step toward automated, user-friendly design of high-performance control sequences.
	\section*{Online content}
	The sequences used in the examples can be found at https://github.com/Jiahui-Chen-quantum/Higher-order-Hamiltonian-engineering.
	\section*{Acknowledgments}
	We thank Alexandre Cooper-Roy for helpful comments on the manuscript.
	This work was supported by the Canadian
	Excellence Research Chairs (CERC) program, the Natural Sciences and Engineering Research Council of Canada
	(NSERC) Discovery program, the Canada First Research
	Excellence Fund (CFREF), and the Innovation for Defence Excellence and Security (IDEaS) Program of the Canadian Department of National Defence.
	\appendix
	\section{Correlated $\mathscr C$-integrals}\label{appen1}
	\begin{figure*}
		\centering 
		\includegraphics[width=0.9\textwidth]{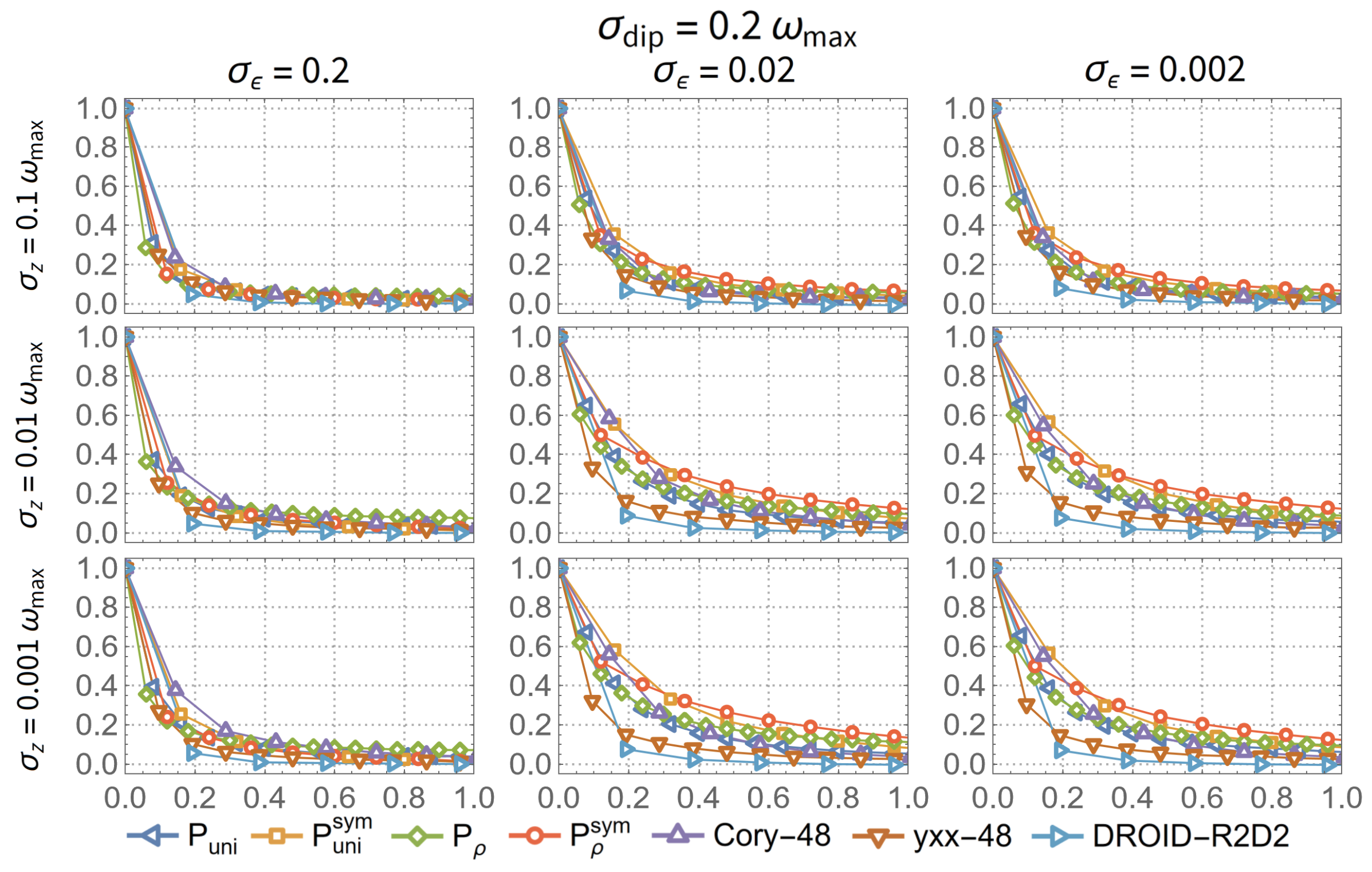}\\
		\caption{Autocorrelation of $\rho_0=\sum_{i=1}^N\x^i$ under different sequences for $\sigma_\text{dip}=0.2\omega_\text{max}=50$ kHz.}
		\label{zhorg1}
	\end{figure*}
	Although the composite $\mathscr C$ space (Eq. \eqref{compc}) of the system studied in Sec. \ref{secexa} is minimal,  
	that of any system is in general not. Following the similar procedure of calculating a $\mathscr C$ space in \cite{chen2025engineering}, the minimal composite $\mathscr C$ space is given by
	\begin{equation}
		\text{span}_{\mathbb R}\left\{\bigoplus_w\ls{\mathcal D_w(g_1)\cdots\mathcal D_w(g_L)|H_\text{pert}^w\rrangle}\middle|g_l\in\mathbf g_\text{pri}, L\ge 0\right\}.
	\end{equation}
	Algorithm \ref{L1} calculates an orthonormal basis of this space.
	\begin{algorithm}[H]
		\caption{Algorithm for finding an orthonormal basis of the minimal composite $\mathscr C$ space.}\label{L1}
		\begin{algorithmic}[1]
			\State $\mathcal B=\lc{\bigoplus_w\mathcal D_w(\text{ad}_{e_i})}_{i=1}^{|\mathbf g_\text{pri}|}$
			\State $|\tilde h_1\rrangle_\text{comp}=\bigoplus_w|H_\text{pert}^w\rrangle_w$
			\State $i_\text{max}\gets1$
			\State $i_\text{new}\gets1$
			\While{$i_\text{new}>0$ and $i_\text{max}\le \sum_w|\mathscr C_w|$}
			\State $t\gets 0$
			\For{$i_\text{max}-i_\text{new}+1\le i\le i_\text{max}$}
			\For{$K\in\mathcal B$}
			\If{$|\tilde h_1\rrangle_\text{comp},\ldots,|\tilde h_{i_\text{max}+t}\rrangle_\text{comp},K|\tilde h_i\rrangle_\text{comp}$ are linearly independent}
			\State $t\gets t+1$
			\State $|\tilde h_{i_\text{max}+t}\rrangle_\text{comp}\gets K|\tilde h_i\rrangle_\text{comp}$
			\EndIf
			\EndFor
			\EndFor
			\State $i_\text{max}\gets i_\text{max}+t$
			\State $i_\text{new}\gets t$
			\EndWhile
			\State Use the Gram-Schmidt process to orthonormalize $(|\tilde h_1\rrangle_\text{comp},\ldots,|\tilde h_{i_\text{max}}\rrangle_\text{comp})$ and \Return the result 
		\end{algorithmic}
	\end{algorithm}
Even when the $\mathscr C_\text{comp}$ is indeed minimal, there can be residual correlation between $\mathscr C-$integrals, in principle, due to constraint in control. For example, it is possible that $\vec c^{(1)}(T_\text{seq})$ do not span $\mathscr C\otimes\mathscr C$. This can be easily checked by calculating the space spanned by $\vec c^{(1)}(T_\text{seq})$ of random sequences. 

If, after sampling many random sequences, a vector in $\mathscr C\otimes\mathscr C$ cannot be expressed as a linear combination of the observed $\vec c^{(1)}(T_{\rm seq})$, it is reasonable in practice to treat the corresponding Hamiltonian as unachievable. Indeed, synthesizing a sequence that realizes a target Hamiltonian is typically much harder than representing its coefficient vector within the span of attainable $\vec c^{(1)}(T_{\rm seq})$.
\section{Performance comparison of decoupling sequences}\label{appen3}
\begin{figure*}[t]
	\centering
	\includegraphics[width=0.9\textwidth]{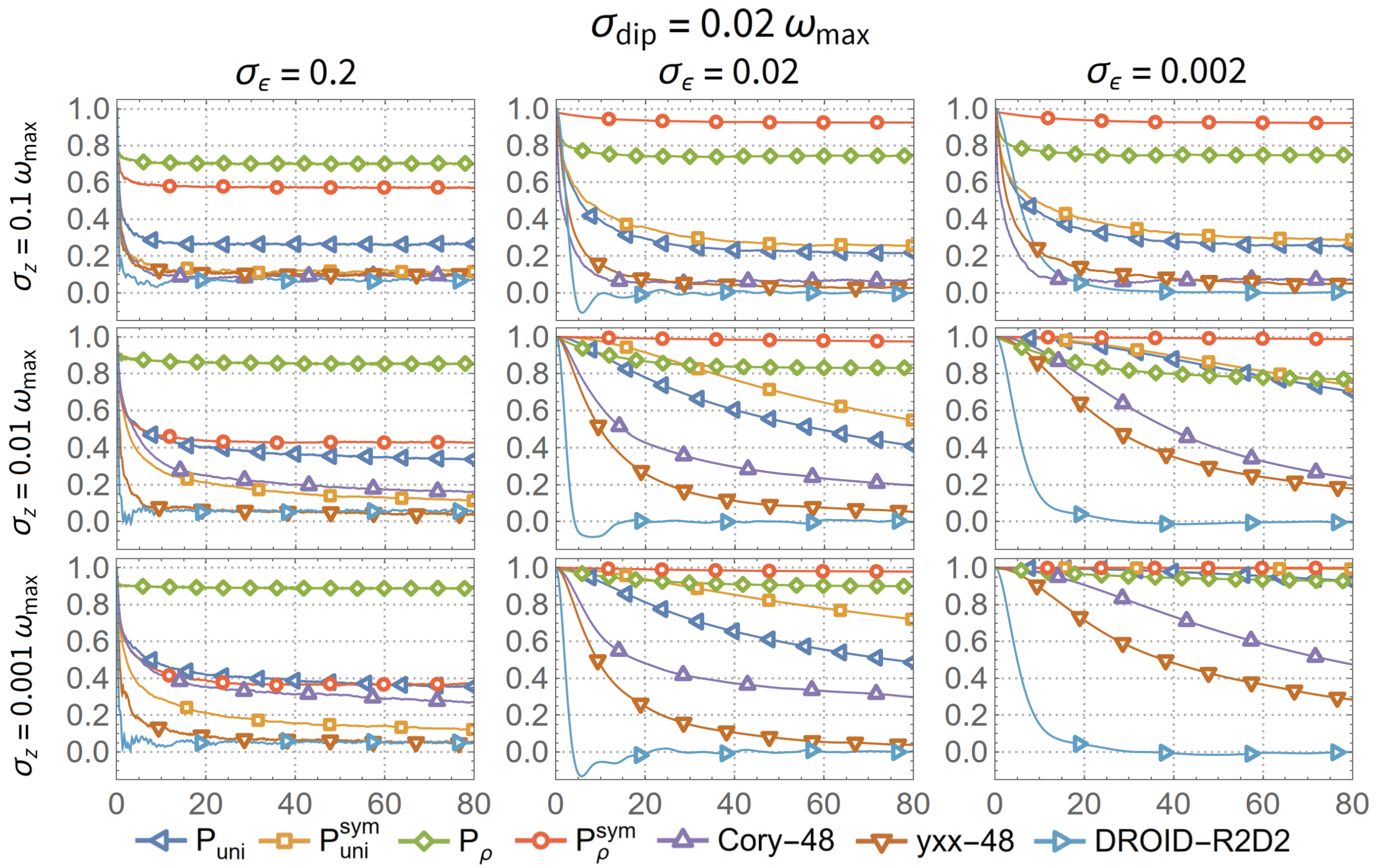}\\
	\caption{Autocorrelation of $\rho_0=\sum_{i=1}^N\x^i$ under different sequences for $\sigma_\text{dip}=0.02\omega_\text{max}=5$ kHz.}
	\label{zhorg2}
\end{figure*}

\begin{figure*}[b]
	\centering
	\includegraphics[width=0.9\textwidth]{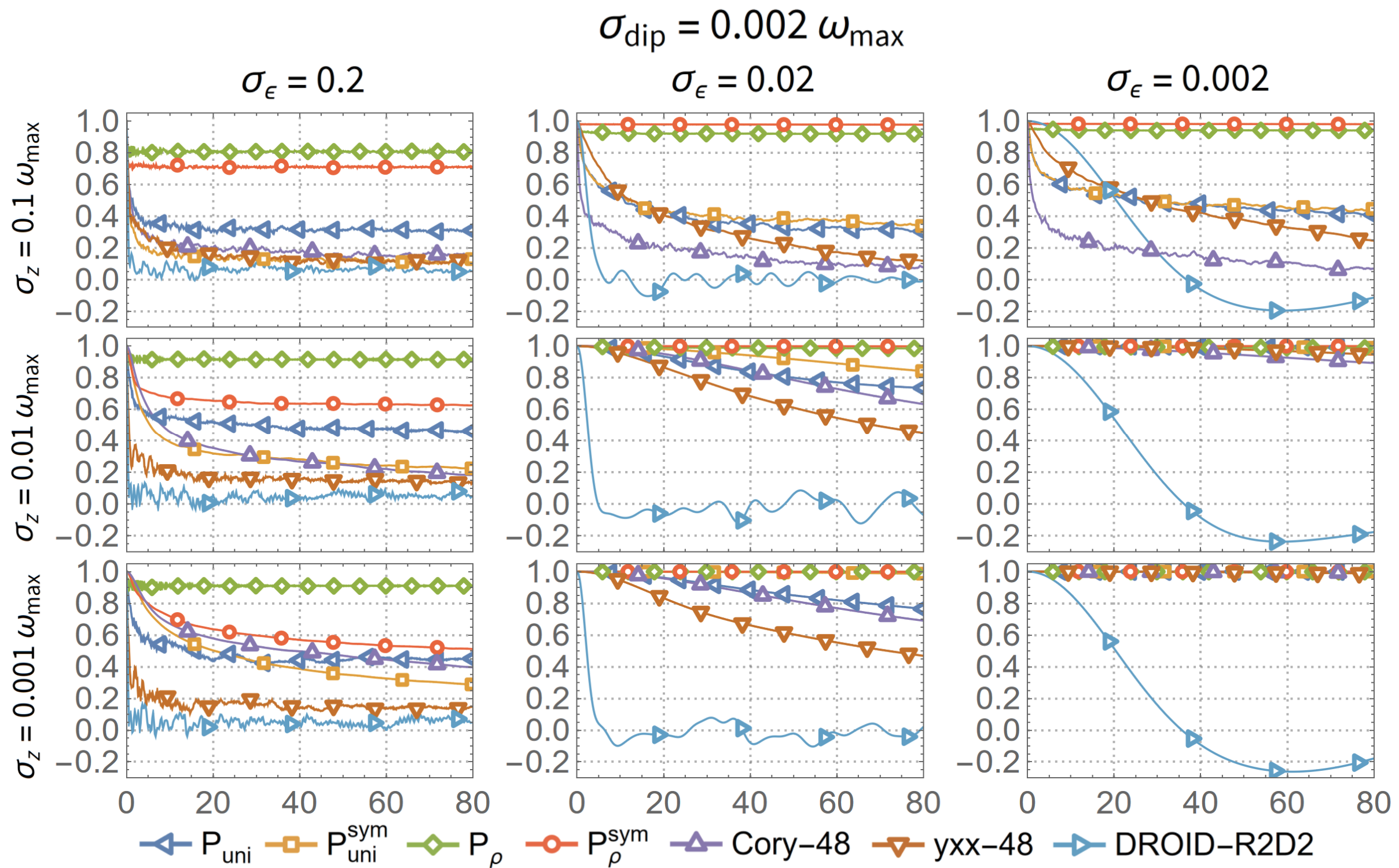}\\
	\caption{Autocorrelation of $\rho_0=\sum_{i=1}^N\x^i$ under different sequences for $\sigma_\text{dip}=0.002\omega_\text{max}=0.5$ kHz.}
	\label{zhorg3}
\end{figure*}

Figs. \ref{zhorg1}, \ref{zhorg2} and \ref{zhorg3} provide performance comparison of the sequences $P_\text{uni}$, $P_\text{uni}^\text{sym}$, $P_\rho$, $P_\rho^\text{sym}$, Cory-48, yxx-48 and DROID-R2D2 at regimes of different error strength. The sequences have the same configurations as in Example \ref{ex1}. The horizontal and vertical axes of the plots are time $t$ (ms) and autocorrelation.

\clearpage
\bibliographystyle{unsrt}
\bibliography{ref1}

\end{document}